\documentclass[iop]{emulateapj}     % two column  

\usepackage{txfonts}
\usepackage{graphicx}

\usepackage{graphicx}
\usepackage{soul}

\usepackage{natbib}

\newcommand{\ks}{\mbox{km~s$^{-1}$}}

\usepackage{natbib}

\begin{document}

\title{Quiet Sun H$\alpha$ transients and corresponding small-scale transition region and coronal heating}

\vskip1.0truecm
\author{
V.~M.~J.~Henriques, D.~Kuridze, M.~Mathioudakis, \& F.~P.~Keenan} 
\affil
{Astrophysics Research Centre, School of Mathematics and Physics, Queen's University Belfast, BT7 1NN, Northern Ireland, UK. \email{v.henriques@qub.ac.uk}}	

\date{Received: 11-20-2015 / Accepted: 02-12-2016 }

%%%%%%%%%%%%%%%%%%%%%%%%%%%%%%%%%%%%%%%%%%%%%%%%%%%%%%%%%%%%%%%%%%%%%%%%%

\begin{abstract}

Rapid Blue- and Red-shifted Excursions (RBEs and RREs) are likely to be the on-disk counterparts of Type II spicules.  Recently, heating signatures from RBEs/RREs have been detected in IRIS slit-jaw images dominated by transition-region lines around network patches. Additionally, signatures of Type II spicules have been observed in AIA diagnostics. The full-disk, ever-present nature of the AIA diagnostics should provide us with sufficient statistics to directly determine how important RBEs and RREs are to the heating of the transition region and corona. We find, with high statistical significance, that at least 11\% of the low-coronal brightenings detected in a quiet-Sun region in 304, can be attributed to either RBEs or RREs as observed in H$\alpha$, and a 6\% match of 171 detected events to RBEs or RREs with very similar statistics for both types of H$\alpha$ features. We took a statistical approach that allows for noisy detections in the coronal channels and provides us with a lower, but statistical significant, bound. Further, we consider matches based on overlapping features in both time and space, and find strong visual indications of further correspondence between coronal events and co-evolving but non-overlapping, RBEs and RREs. %We find evidence for jet generating sources that a  source of co-spatial and co-temporal 304 and H$\alpha$ signatures where the H\alpha. % that do not always have a signature in H$\halpha$ suggesting low-corona heating at

\end{abstract}

%%%%%%%%%%%%%%%%%%%%%%%%%%%%%%%%%%%%%%%%%%%%%%%%%%%%%%%%%%%%%%%%%%%%%%%%%
\section{Introduction}

Recent high-resolution, multi-instrumental and multi-wavelength observations of chromospheric transients have re-introduced the question of how much do these affect the outer layers in terms of heating and mass loading. 

RBEs (Rapid Blue-shifted Excursions) and RREs (Rapid Red-shifted Excursions) are small-scale transients observed in either the blue or the red far-wings of  chromospheric lines \citep{2008ApJ...679L.167L, 2009ApJ...705..272R, 2012ApJ...752..108S, 2013ApJ...769...44S, 2015ApJ...799..219D}. The measured properties of RBEs are similar to those reported for type II spicules, first reported in Ca~{\sc{ii}}~H \citep{2008ApJ...679L.167L, 2009ApJ...705..272R}, and are therefore thought to be the H$\alpha$ counterparts of these structures. In this work, the term Rapid Excursions (REs) will be used to refer to both RBEs and RREs.

\cite{2011Sci...331...55D} reported that RBEs have signatures in transition region (TR) lines. The observed FOV was that of an active region displaying striking coronal loops. 
Furthermore, \cite{2015ApJ...799L...3R} found a spatio-temporal match between H$\alpha$ REs and features in the IRIS TR passbands. In particular, they found evidences that there are spectral signatures for REs in IRIS C~{\sc{ii}} 1335 and 1336 {\AA} and Si~{\sc{iv}} 1394 and 1403 {\AA} spectral lines. 
However, in the lower activity network regions, they found only weak signals in the C~{\sc{ii}} and Si~{\sc{iv}} spectra with none apparent in the corresponding slit-jaw imaging channels (i.e. 133 and 140 nm) that can be traced back to RBEs. 
This is despite an abundance of RBE detections along the slit.  Furthermore, they observed a diffuse halo around the stronger network elements and network-jet like signals that match the RBEs in the spectra.  
The IRIS detections show that REs have a heating impact up to 80~kK (Si~{\sc{iv}}), and at least in areas as quiet as network regions.

\cite{2014ApJ...792L..15P} studied Type II spicules at the limb using Hinode, IRIS, and SDO/AIA. They find matching heating signals in AIA~He~{\sc{ii}}~30.4~nm with a small spatial displacement  from the chromospheric Ca~{\sc{ii}}~H signals (i.e. continuing the path of Ca~{\sc{ii}}~H spicules, see their Fig. 3). This offset indicates that spicules are being heated as they travel upwards, disappearing from the cooler channels, with signatures appearing in the hotter channels with matching signal found in lines as hot as Si IV 140~nm.  In a similar set of data, the temporal evolution of type II spicules was studied by \cite{2015ApJ...806..170S} who find stronger evidence for progressive heating. The He~{\sc{ii}}~30.4~nm detections clearly shows that at least some type II spicules reach temperatures as high as 200~kK.

\cite{2014Sci...346A.315T} detected large-scale jets with speeds of 80-250~\ks, seen above network regions using IRIS slit-jaw images in the 133~nm~Si{\sc{iv}} and 140~nm~C{\sc{ii}}  passbands. The latter were observed in bundles, on disk locations close to the limb above network bright-points. \cite{2014Sci...346A.315T} argue that these "network jets'', due to the observed temperature and velocity, likely do not fall back on the solar surface but instead contribute to the heating of the outer atmosphere and mass of the solar wind. 

\cite{2015ApJ...807...71P} used AIA observations at the edge of a coronal hole at the limb, to study the property of propagating disturbances that they identify as "jetlets". They found that these jetlets have lower velocities than the IRIS network-jets reported by \cite{2014Sci...346A.315T}, a difference that they attribute  to the lower activity of the region. They also found a correspondence between some AIA signatures and network-jets in slit-jaws images from IRIS, as well as some events without any correspondence. Similarly, the IRIS detections of \cite{2015ApJ...799L...3R} also appear to be lower in velocity than those of \cite{2014Sci...346A.315T}. However,  \cite{2015ApJ...799L...3R}  find a clear relation between some of their REs and the TR network-jets of \cite{2014Sci...346A.315T}.

In active regions, jets with properties consistent with RBE's have been observed to be important for the heating of coronal loops \citep{2009ApJ...701L...1D}. More recently, based on in AIA observations alone, further evidence has been put forward to suggest that "plume" or "jetlet" transients are the main source of heating in coronal loops \citep{2014ApJ...787..118R}. 

Models of the physical processes that could lead to chromospheric heating from transients can be found as early as 1982 \citep{1982ApJ...255..743A}. 
Evidence that this energy, deposited in the chromosphere, can make it's way into the corona has also emerged from simulations \citep{2005ApJ...618.1031G,2007ApJ...656..577A}. 

Recently,  \cite{2011AIPC.1356..106Z} proposed that the highly dynamical chromospheric jets could be unstable due to the Kelvin-Helmholtz instability
produced by velocity discontinuities between the surface of the jet and surrounding media. This could be responsible for the rapid heating, and hence the observed fast disappearance of REs/type II spicules in the chromosphere \citep{2015ApJ...802...26K}.  

While the detailed modeling of solar atmospheric heating based on transients events remains an open question, it is observationally clear that jet-like transients have an impact of at least the lower corona and TR. What is not observationally clear is how big of an impact these transients have, especially for the most prominent regions on the Sun such as coronal holes and quiet Sun regions that do not display prominent loop structures.  

In this paper we try to answer the question of how important RBEs and RREs are to the lower corona and transition region. For this we look at a typical quiet Sun region. Our approach is to find coronal and transition region heating signatures and relate those signatures to transients observed in the chromosphere. We use AIA data due to its continuous sampling of the Sun, due to the link with the previous results and, most importantly, due to the potential for large statistical significance. SST data are used for the chromospheric observations providing high-spectral, spatial and temporal resolution in H$\alpha$. Limb observations are highly revealing but it is difficult to derive statistics from them due  to multiple overlapping  structures along the observer's line-of-sight. At disk centre this limitation does not exist and we use such advantage to automatically match structures in this work.

%%%%%%%%%%%%%%%%%%%%%%%%%%%%%%%%%%%%%%%%%%%%%%%%%%%%%%%%%%%%%%%%
\section{Observations and Data Processing}
\label{sect:setup}

\begin{figure*}[!htb]
  \centering
\resizebox{5.5cm}{!}{\includegraphics[clip=true]{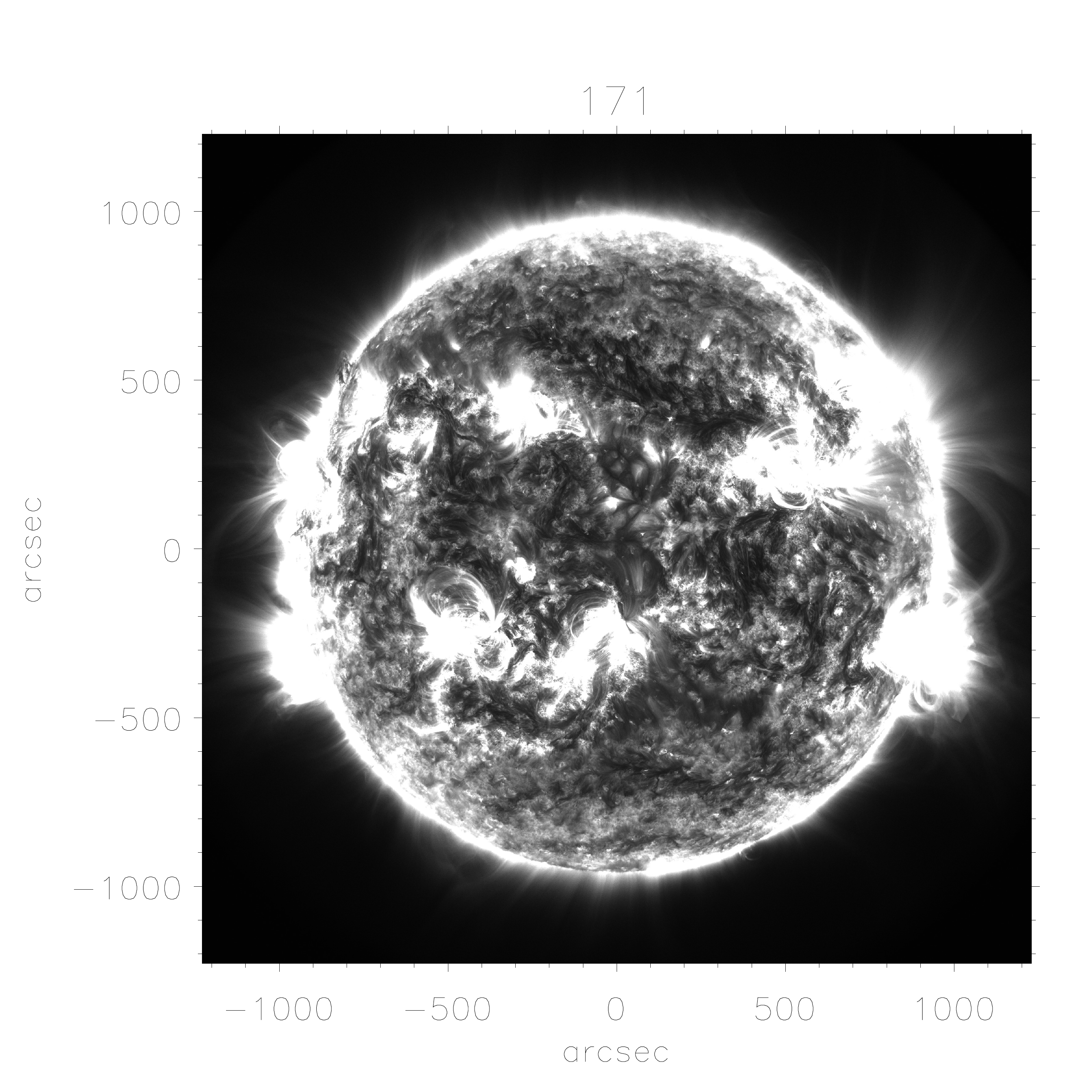}}
\resizebox{5.5cm}{!}{\includegraphics[clip=true]{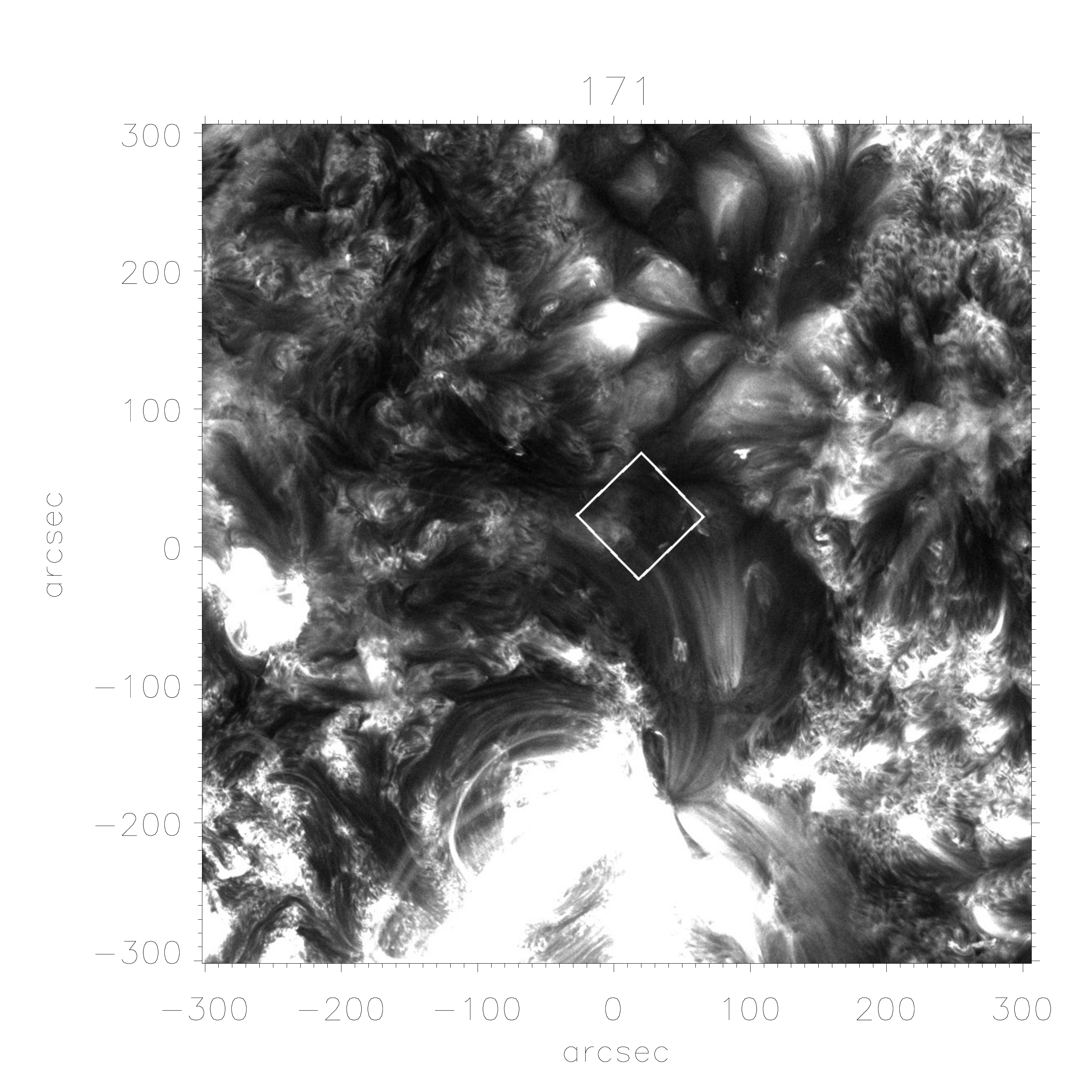}}
\resizebox{5.5cm}{!}{\includegraphics[clip=true]{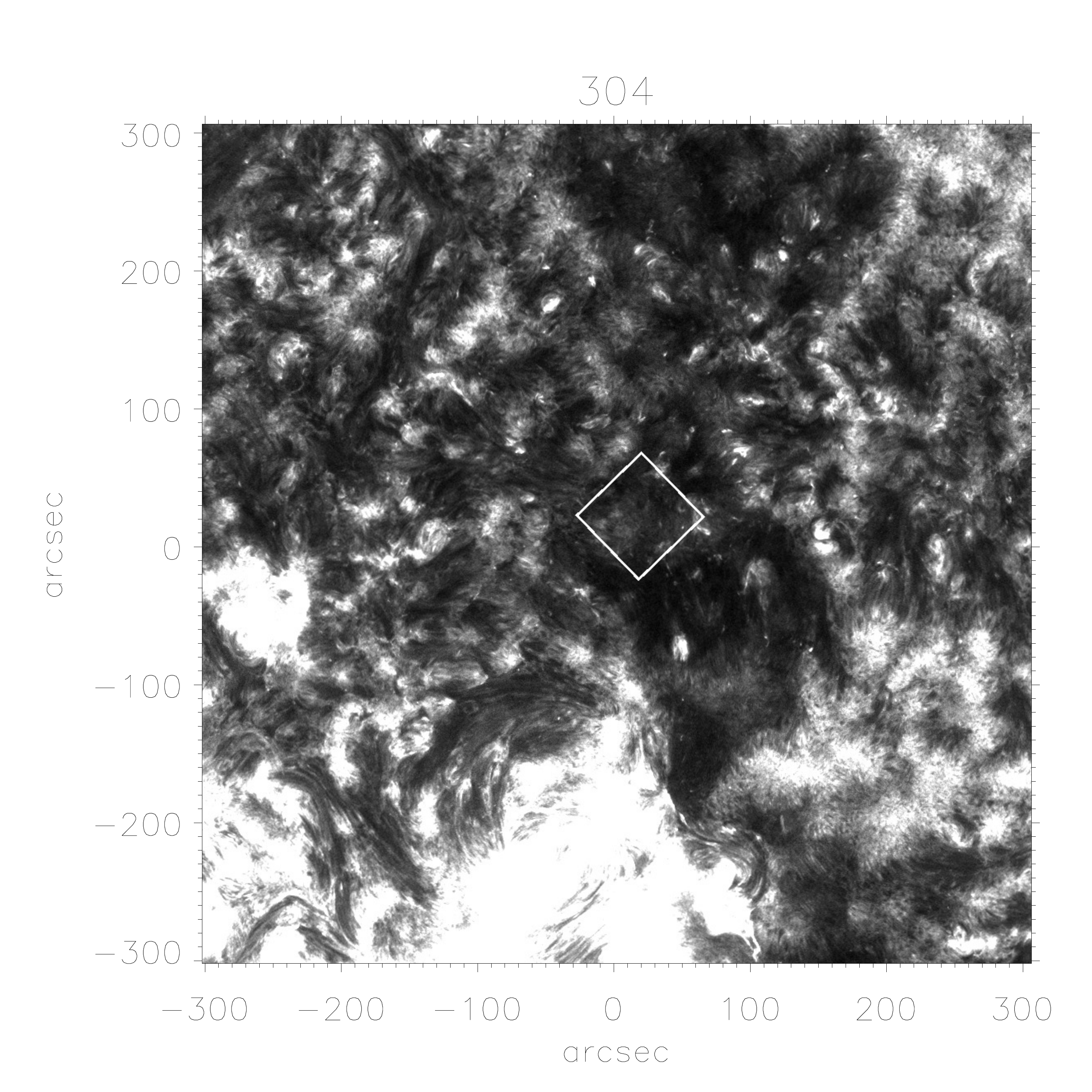}}
  \caption{\footnotesize  Left panel:  A full sun image in Fe~{\sc{ix}}~171~\AA . Center: A zoom in showing the region under investigation and context in 171 \AA . Right: The same region in He~{\sc{ii}}~304~\AA\ . All panels have been scaled to a 10\% saturation level for better visibility of the quiescent region under investigation. The central square denotes the FOV observed in H$\alpha$ with the SST.}
\label{quiet}
\end{figure*}

The observations presented in this paper were obtained between 09:06 and 09:35 UT on 2013 May 3, and the target was a very quiet region located at disk centre (Fig.\ref{quiet}). This compares to more active areas of the Sun presented in the earlier work of \cite{2011Sci...331...55D}. Two rosettes are visible, anchored in two sets of photospheric bright points and show remarkably little evolution throughout the time-series. Both rosettes show activity in H$\alpha$ with the upper rosette showing torsional motions in alternating directions. Some of these rotational motions  translate to transverse displacements of REs which have been analysed by \cite{2015ApJ...802...26K}.  This paper describes the ground-based observations and reduction, and hence here we only present a brief summary.  %\cite{2015ApJ...802...26K}

We used the CRisp Imaging SpectroPolarimeter \citep[CRISP;][]{2006A&A...447.1111S,2008ApJ...689L..69S} instrument,
at the 1-m Swedish Solar Telescope \citep[SST;][]{2003SPIE.4853..341S}. Adaptive optics, including an 85-electrode, were used \citep[an upgrade of the system described in][]{2003SPIE.4853..370S}. 
All data were reconstructed with Multi-Object Multi-Frame Blind Deconvolution \citep[MOMFBD;][]{2002SPIE.4792..146L,2005SoPh..228..191V}, using 51~Karhunen-Lo\`{e}ve modes sorted by order of atmospheric significance and $88\times88$ pixel subfields. 

A prototype of the data reduction pipeline published by \cite{2015A&A...573A..40D} was used before and after MOMFBD. This includes the method described by \cite{2012A&A...548A.114H} for alignment and destretching as in \cite{1994ApJ...430..413S}. 

\begin{figure*}[t]
\begin{center}
\includegraphics[width=18cm]{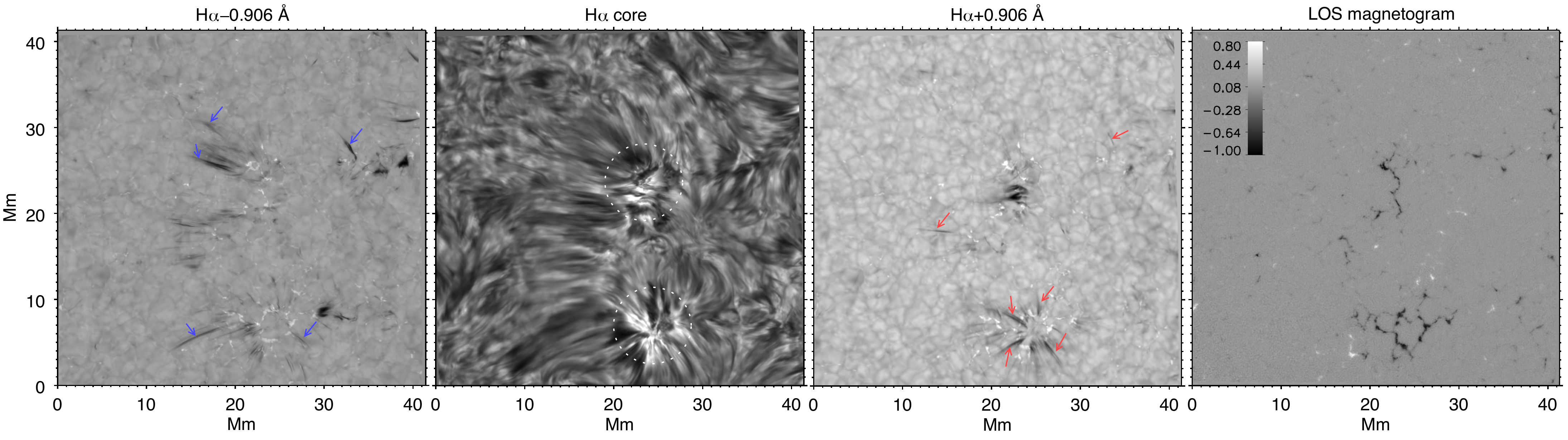}
\end{center}
\caption{H$\alpha$ core and $\pm$0.906 {\AA}  images
together with the photospheric, line-of-sight magnetogram obtained from Fe~{\sc{i}}~6302 {\AA} Stokes {\it V} profiles.
Red and blue arrows indicate typical RREs and RBEs observed in the H$\alpha$ line wings.
The rosette regions where most of the RBEs/RREs are detected are highlighted with white dotted circles.
The images show that the footpoints of the RBEs/RREs correspond to photospheric bright points and strong magnetic flux concentrations.  
The color scale in the magnetogram indicates the magnetic field strength in kilogauss. Figure reproduced from \cite{2015ApJ...802...26K} for context. }
\label{rbe_context}
\end{figure*}

The spatial sampling is 0$''$.0592 pixel$^{-1}$, with the spatial resolution reaching up to 0$''$.16 in H$\alpha$ over the field-of-view (FOV) of 41$\times$41 Mm (Fig.\ref{rbe_context}).
The H$\alpha$ line scan consists of 7 positions (-0.906, -0.543, -0.362, 0.000, 0.362, 0.543, +0.906 {\AA} from line core) corresponding to a range of -41 to +41~\ks in velocity, and the temporal cadence of the full spectral scan was 1.34 s.

We also use imaging data from the Atmospheric Imaging Assembly (AIA), on board of the Solar Dynamics Observatory \citep{2012SoPh..275...17L} in the He~{\sc{ii}}~304\AA\ and 171\AA\ pass-bands (henceforth 304 and 171, respectively). The data were downloaded and processed to level 1.5 using Solarsoft  \citep{2000eaa..bookE3390F} which includes co-alignment between the AIA channels up to 2 \arcsec. We then aligned, de-rotated, and re-sampled the AIA data to the SST H$\alpha$ images. All analysis is performed at the SST pixel-size for all data. Alignment using manual control points, Solarsoft and standard IDL routines were used. The AIA channels were co-aligned to better than one AIA pixel (0.6 \arcsec) using manual comparison with blinking after the initial alignment and resampling to the SST scale.  Sub-pixel shifting with cubic  interpolation was used for all channels during the re-alignment procedures.

%%%%%%%%%%%%%%%%%%%%%%%%%%%%%%%%%%%%%%%%%%%%%%%%%%%%%%%%%%%%%%%%
\section{Data Analysis}
\label{analysis}

Our goal is to create data cubes consisting of brightening detections from the 304 and 171 channels, and data cubes consisting of blue wing, red wing, RBE, and RRE detections from H$\alpha$ for comparison. Running-difference cubes are computed for all observations (AIA and SST) comprising of the difference between the mean of the three central adjacent frames (30 seconds), and the mean of the two frames located 20 seconds before and 20 seconds after the central frames. This is aimed high sensitivity to transients with life-times around those of RE (50 seconds). Previous detection approaches have used differentiation with H$\alpha$ frames close in time but in the photospheric wings \citep[with the SST][]{2012ApJ...752..108S} or differentiation with the previous frame in a time-series of binned frames \citep[in AIA and Hinode][]{2011Sci...331...55D}. Our approach in this step is most similar to the latter. In all aforementioned works as well as this, wavelengths just over 40 \ks from  H$\alpha$ core are used.

Binary maps are computed to locate the events by using thresholding and basic morphological operations. For 304 and 171 a threshold of the mean plus 1.5 standard deviations is applied. Then a second mask, the hit mask, selects the binary detections with at least one pixel above 1.9 standard deviations. For H$\alpha$ a single mask with a threshold of the mean plus 1.1 standard deviations is used. 

An erode, a dilate and another erode are performed on all H$\alpha$ binary masks using disks of 1, 4, and 1 pixel diameter, respectively. This was intended to work both as a morphological close and noise removal at the smallest scales.  For 304 and 171 one single dilate with a disk of one SST pixel (0\farcs058) was used to close regions adjacent at the vertices that would otherwise be disconnected. Each contiguous region was marked as one individual detection using a label operation. Standard IDL routines were used for the morphological operators and labelling which should be available in any standard computer vision package. The labelling step marks structures in the three-dimensional cubes thus tracing structures in space and time. Relating with previous automated-detection work, the morphological operations and labelling would be analogous to the distance thresholds used in \cite{2012ApJ...752..108S} to determine that two nearby structures in time and space are the same. Unlike \cite{2012ApJ...752..108S}, we don't select for elongation as we often find events originating in groups, occasionally forming wide detection blobs that would be completely discarded. We also find small blob-like RBEs with little elongation, but very high contrast and a clear propagation direction, in the difference maps that would have been discarded had an elongation threshold been considered.  

Finally, all detections below a size of 500 SST pixels in time and space (that is, in the three-dimensional data-cube) were removed. We further separate the H$\alpha$ detections in "blue" and RBEs as well as "red" and RREs. The difference between the RE maps and the respective wings maps is that the RE maps were zeroed where a signal was also picked up in the opposite wing.  

\begin{figure*}[!htb]
  \centering
\resizebox{18.5cm}{!}{\includegraphics[clip=true,trim=0.3cm 0 0.3cm 0]{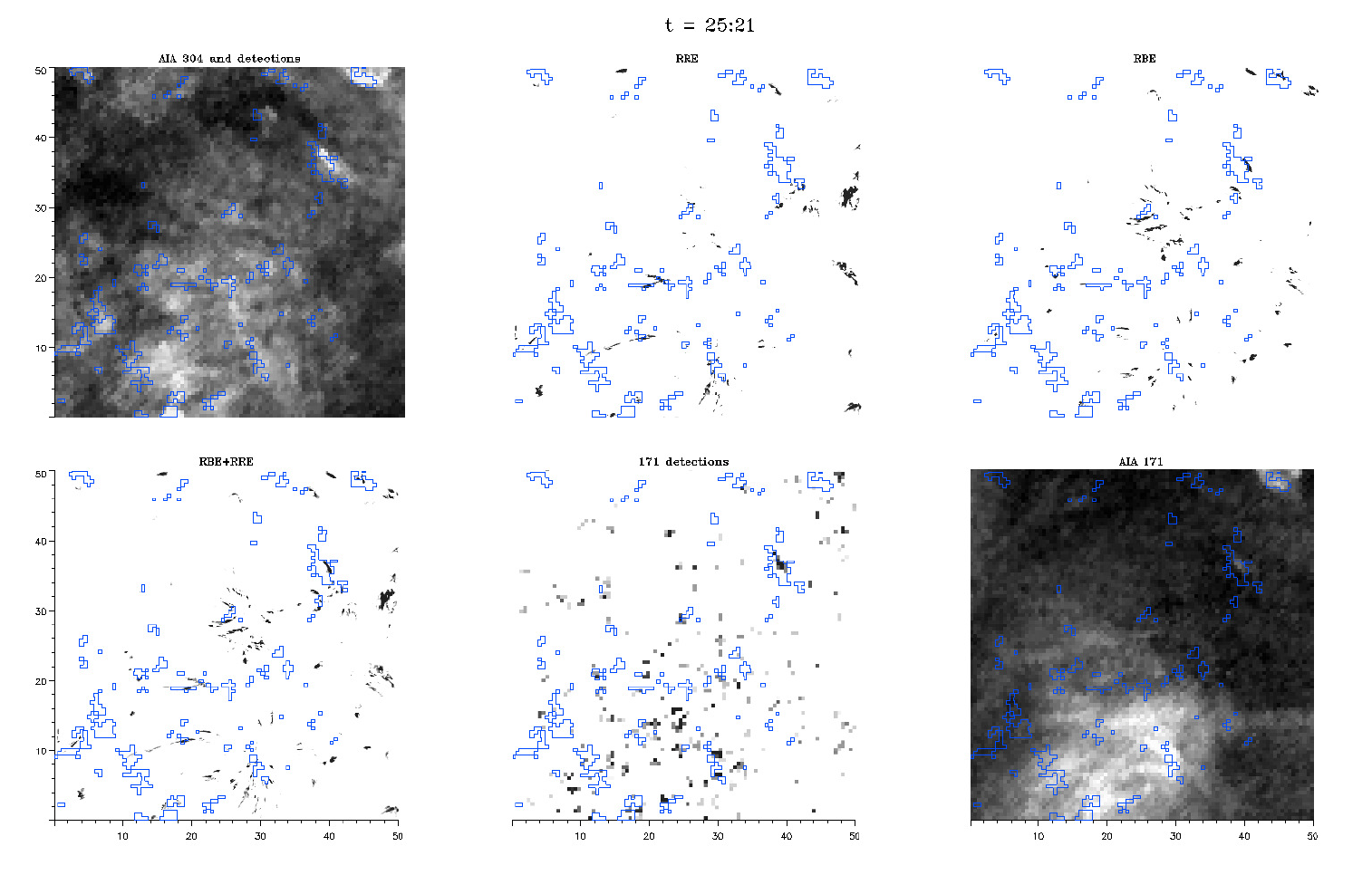}}
  \caption{\footnotesize A single frame showing a snapshot of the analysed data where the contours display the 304 detected heating events over all maps. Top left: 304 image. Centre top: RRE detections. Top right: RBE detections. Bottom left: all REs. Centre bottom: 171 detections. Bottom right:  171 imaging. Associated time-series available. 304 and 171 images clipped at 5\% on both intensity extremes for display purposes. Units are in \arcsec . }
\label{movie1}
\end{figure*}

In Fig.\ref{movie1}, and the associated time-series, the contours of the detection maps obtained from 304 are shown plotted over 304 and 171 as well as  the detection maps of the RBEs, RREs and 171.  
Statistics of the matches between the different detection maps are shown in Table~\ref{table:stats2} for matches with 304, and Table~\ref{table:171} for matches with 171. We consider a detection to have a match with another wavelength map if it overlaps by at least 100 SST pixels, or $0.43 \mathrm{arcsec}^2$, in time and space with detections from that other map. If 100 pixels is less than 5\% the size of the structure being matched, times the filling factor of the map it is being matched against, then the latter value is the threshold of overlap above which a match is considered. This is done to prevent the largest structures from having a match with no statistical significance. These thresholds are also computed in this way when finding the individual probabilities of a match ($P_i$) due to randomness in Section~\ref{statistics}. The reasoning behind the 5\% value is also explained in Section.~\ref{statistics}.  

The selection of the threshold levels was made iteratively such that obvious events are always detected but separate from nearby events while, at the same time, keeping the total amount of pixels in a cube small. However, small changes do not seem to generate significantly different patterns. While the number of removed small detections, detected noise (in the AIA case), and the size of the detected regions will change with different thresholds and filtering, our main aim is to make sure that the statistical significance of the results is high. For the thresholding of AIA maps this mainly means that the total amount of detected pixels should remain a small fraction of the whole FOV, and that the overlap criteria remains comparatively high (discussed in Section~\ref{statistics}). %We study the statistical significance of this match parameter and our results in Section~\ref{statistics}.   %

\begin{table*}
\caption{Statistics of matched 304 detections to other channels.}             
\centering          
\begin{tabular}{c l c c c c c}     % 7 columns 
\hline\hline    
\footnotesize Quantity & 304 matches with & Lower bound at 99\% confidence & Quantity matched with 304 & Lower bound at 99\% Confidence  \\ 
     H$\alpha$ RBE & 259 (11\%) & 9\% & 342 (35\%) & 31\% \\
     H$\alpha$ RRE & 214 (9\%) & 7\% & 285 (32\%) & 28\% \\
     H$\alpha$ Blue & 269 (11\%) & 9\% & 317 (36\%) & 32\% \\
     H$\alpha$ Red & 223 (9\%) & 7\% & 274 (36\%) & 31\% \\
     171 & 874 (37\%) & 34\% & 479 (18\%) & 16\% \\
     any RE  & 388 (14\%) & 13\% & - & - \\
\end{tabular}
\tablecomments{\footnotesize Statistics of the matched detected regions. First  column: number and respective percentage of 304 detections that were matched with the structures of the left. Second  column: lower confidence bound at 99\% using Gaussian statistics for the percentage in the first column.Third  column: number and respective percentage of detections from the quantity on left that was matched with 304. Final column: lower confidence bound at 99\% using Gaussian statistics for the percentage in the previous column.  \label{table:stats2}}
\end{table*}

\begin{table*}
\caption{Statistics of matched 171 detections to other channels.}             
\centering          
\begin{tabular}{c l c c c c c}     % 7 columns 
\hline\hline    
\footnotesize Quantity & 171 matches with & Lower bound at 99\% confidence & Quantity matched with 171 & Lower bound at 99\% Confidence  \\ 
     H$\alpha$ RBE & 171 (6\%) & 5\% & 383 (39\%) & 35\% \\
     H$\alpha$ RRE & 130 (5\%) & 3\% & 342 (39\%) & 35\% \\
     any RE  & 260 (10\%) & 8\% & - & - \\
\end{tabular}
\tablecomments{\footnotesize Statistics of matched detected regions. First  column: number and respective percentage of 171 detections that were matched with the structures of the left. Second  column: lower confidence bound at 99\% using Gaussian statistics for the percentage in the first column.Third  column: number and respective percentage of detections from the quantity on left that was matched with 171. Final column: lower confidence bound at 99\% using Gaussian statistics for the percentage in the previous column.  \label{table:171}}
\end{table*}

\begin{table*}
\caption{Statistics for the Null hypothesis.}             
\centering          
\begin{tabular}{c l c c c c c}     % 7 columns 
\hline\hline   
\footnotesize Hypothesis (H$_0$) & H$_0$ Expected value ($\mu$)  & p-value & Chernoff bound (log$_{10}$)  \\   
    304 random match with RBE  & 55 (2\%) & 32 & -85 \\
     304 random match with RRE   & 41 (1\%) & 31 & -77 \\
     304 random match with any RE & 91 (3\%) & 36 & -115 \\
     171 random match with RBE & 40 (1\%) & 24 & -50 \\
     171 random match with RRE & 28 (1\%) & 22 & -41 \\
     171 random match with any RE & 96 (3\%) & 19 & -41 \\
     RBEs random match with 304  &  94 (10\%) & 30 & -84 \\
     RREs random match with 304   & 69 (8\%) & 30 & -82 \\
     RBEs random match with 171 & 136 (14\%) & 25 & -1.85 \\
     RREs random match with 171 & 108 (12\%) & 26 & -70 \\
\end{tabular}
\tablecomments{\footnotesize Statistics for the Null hypothesis. First column: number of matches that are expected from pure random coincidence and respective percentage. Second column: p-value from a Z-test statistics (Gaussian). Third column: upper Chernoff bound giving the probability that the observed match listed in Table~\ref{table:stats2} and Table~\ref{table:171} could have been due to pure random coincidence.\label{table:chernoff}}
\end{table*}

%%%%%%%%%%%%%%%%%%%%%%%%%%%%%%%%%%%%%%%%%%%%%%%%%%%%%%%%%%%%%%%%
\subsection{Statistical testing}
\label{statistics}

We choose as the null hypothesis (H$_0$) that the overlaps of the transient events across two passbands are due to pure coincidence or matches with noise, and assume that the transient detections can occur randomly at any point in time and space (as noisy detections would). Every single labelled detection is taken, with its properties of a size in time and space, and the probability of that region satisfying our match criteria against the full pixel-space of the second passband, is computed. The probability of a region being matched then follows a Bernoulli distribution with the parameters being our pixel overlap criterion (i.e, a minimum of 100 matching pixels or 100 Bernoulli successes), the size of the region (the number of Bernoulli trials), and the probability of each trial being a success. The latter is given by the total filling-factor of the detections in the second passband (i.e. approximately 1.3\% for RBEs and 1\% for RREs). We compute this probability for every single detection ($P_i$).  

With the individual probability of a match for each region known, finding the total probability of a certain number of matches between  becomes a classic Poisson-trial problem, where our random variable X is the sum of the outcomes of the several Bernoulli variables with different probabilities. Under certain circumstances, the probability distribution of X can be approximated by a Gaussian with $E[\hat{X}]$ and $\sigma[\hat{X}]$ as parameters.  However, an approximation by a Gaussian may not be justifiable for this problem since the probabilities of each trial fluctuate between extremely low values, and values of the order of one due to the low filling factors of the H$\alpha$ maps combined with the very different sizes of each 304 and 171 detection. Thus we chose to compute a stronger test statistic, without having to do any approximation or even having to know the exact distribution of X, by using Chernoff bounds. We use the upper Chernoff bound \citep[see for e.g.][]{probbook} which provides an exact upper limit on the probability that our number of matches (or number of Bernoulli successes) are higher than our measured value (k) for the null hypothesis:

   \begin{equation}
       P\left(X>k\right) < \mathrm{e}^{(k-E[X])} \left(\frac{E[X]}{k}\right)^k ,
      \label{eq:1}
   \end{equation} 

where E[X] is the expected value for the number of successes given by $E[X]=\sum P_i$. This bound yields a probability lower than $10^{-50}$ for H$_0$ for every H$_0$ except RBEs to 171 which yields $P[X > k] < 0.014$ (see Table \ref{table:chernoff}), which definitely excludes the null hypothesis of the matches occurring at random for every case. This value is striking, but not surprising when we consider that we are dealing with small filling factors while requiring a broad overlap match (minimum of 100 pixels). Nor is it surprising considering the actual number of matched structures and the degree of visual match presented in our figures, and associated time-series, when compared to what would be expected from chance alone. 

For completeness and familiarity, we also compute the p-value for a Z-test statistic, which assumes a Gaussian distribution for X as described above. We obtain a p-values of two digits for every statistic. This means that our match statistics are tens of standard deviations away from the expected value of the null hypothesis assuming Gaussian statistics. We can therefore confidently reject the null hypothesis when assuming Gaussian statistics to much better than 99\% for all cases.  

Note that if our detection algorithms were picking mainly noise, then our test statistics would be necessarily close to those of the null hypothesis, at least within the confidence interval. One could still argue, disregarding the very different band-passes and instruments used, that systematic errors in the detection method are generating noise at the same locations in both H$\alpha$ and 304. While the 304 data are undoubtedly noisy, the H$\alpha$ detections follow very obvious, very high-contrast features in both H$\alpha$ wings and are therefore easier to detect. Additionally, from the time-series,  the 304 detections do seem to follow obvious intensity changes over the RE time-scales. 

Regarding the dependency of the test statistics on our specific choice of the match criteria, we note that Chernoff bounds drop exponentially with the number of successes (see Eq.\ref{eq:1}). Our criteria for matching structures do not decrease as fast due to the large number of detections that clearly overlap with those of H$\alpha$ over thousands of pixels in space and time, as may be seen in the figures and associated time-series discussed in Section \ref{discussion}. Extensive testing with different overlap overlap criteria was done that confirms these dependencies. Further, these tests were used to select the final overlap criterium itself. We essentially aimed at selection criteria that would be undisputably significant but not unecessarily harsh leading to underestimated matches. This was achieved by simply selecting criteria that led to expected values for the null hypothesis ($E[X]$ for $H_0$), of about 1\% for the AIA to SST cases. As mentioned in Section.~\ref{analysis}, this ended up being a minimum of 100 pixel overlap or 5\% the size of the structure being matched, times the filling factor of the map the structure is being matched against, whichever is higher. The 5\% threshold mostly applied to unusually large 304 structures visible shooting from the the lower rosette (e.g. Fig.\ref{lower}).

Everything considered, we successfully reject our null hypothesis of the computed matches in Tables~\ref{table:stats2} and ~\ref{table:171} being due to the random coincidences across all maps, for our selected criteria and using non distribution-dependent hypothesis testing. 

%%%%%%%%%%%%%%%%%%%%%%
\section{Discussion}
\label{discussion}
%\subsection{Visual Analysis}

\subsection{Statistics of matching coronal signatures with REs}

We find that 11\% of the 304 detected brightenings match RBEs, with a 9\% match between 304 and RREs. The $99\%$ lower confidence levels using Gaussian statistics are 9\% and 7\%, respectively (see Table.~\ref{table:stats2}). Computing the match between 304 and any RE we obtain a value of 14\%, which is not significantly higher than the RBE match value if one considers the respective increase in the expected value from noisy matches as measured by H$_0$ (Table~\ref{table:chernoff}). RBEs and RREs are often in the vicinity of each-other so it is not surprising that the match with REs is not a sum of the matches from RBEs and RREs.

For 171, about 6\% of the detections can be traced back to RBEs and 5\% to RREs (see Table.~\ref{table:171}). While these values may appear too low to claim that a relation exists, they are highly significant as we have completely invalidated the null hypothesis of this result being due to chance by using distribution independent Chernoff bounds, and with expected matches due to chance of just $1\%$ for both 171 to RBEs and 171 to RREs. To our knowledge, it is the first time that such a statistical match has been demonstrated between chromospheric jet-like features and coronal events. Further, to our knowledge, it is the first time an estimate for the relation between signatures in the corona in the 0.8~MK region and quiet-Sun chromospheric events is proposed. Higher match values would be obtained by selecting lower detection thresholds for the 171 difference maps or by lowering the match-by-overlap criteria. However, any of the latter would translate into a higher expected value from random matches in our robust null-hypothesis testing. To put this into other words: the main effect of the amount of detections due to noise in 304 and 171 is to weigh on the selection of our match criteria following the procedure discussed in Section~\ref{statistics}, and thus lower the computed number of detections, rather than lead to a lower significance of the result. Thus, we speculate that future higher-resolution, higher-aperture observations of the corona will reduce the noise of automated detection algorithms and thus lead to higher statistical estimates of the 304/171 to chromospheric RE matches. 

Finally, we computed the match percentage of 171 with any type of RE for which we obtain a value of 10\%. Similarly to the 304 case, this is not significantly higher than the 171 to RBE match percentage (6\%) if one simply subtracts the respective increase in the expected value from noisy matches (3\%).

\subsection{Statistics of REs and RE to corona matches}

We find that about 25\% of all RE features can be matched with both 304 and 171 signatures if one removes the expected values from chance alone from the measured matches (see Tables 1 through 3). 

For the H$\alpha$ detection maps the contrast is very high and thus, unlike for 304 and 171, the absolute number of events detected is a relevant result. We have detected 974 RBEs and 864 RREs. A few of these structures may be part of the same feature but disconnected due to the common close proximity of RBEs and RREs, leading to a simple blue-wing detection to be split into two RBEs because half way there is also overlapping red-wing signal. The blue wing and red wing features detected were 867 and 759 respectively which we believe is a more reliable number for the actual amount of transients observed. We generally find little difference between our wing detection-maps and the RE maps. Finally, this close proximity of RBEs and RREs, more visible in the lower rosette (Fig.\ref{lower} and time-series associated with Fig.~\ref{movie1}), may be further evidence of the twist observed by \cite{2012ApJ...752L..12D} and \cite{2014Sci...346D.315D}. It is also further evidence (together with the work of \cite{2015ApJ...802...26K}) that RBEs and RREs are likely aspects of the same phenomena. 

The total volume in time and space of the RBEs and RREs, divided over the total three-dimensional cube volume, gives a filling factor of 1.3\% and 1\% respectively. This value should be useful for future studies on the importance of these structures for the quiet-Sun dynamics.
The estimated filling factor is consistent with the filling factor of type II spicules obtained by \cite{2010ApJ...719..469J}. 
They assumed type II spicules, observed with HINODE in Ca II H filter, to be randomly distributed along the boundaries of 
circular supergranules with each spicule having a diameter of 0.1 Mm, and obtained an area filling factor of 1.5 $\%$ \citep{2010ApJ...719..469J}. 
For comparision, an estimate for the filling factor of type I spicules in the quiet Sun chromosphere is 4-5$\%$  \citep{2003PNAOJ...7....1M,2012JGRA..11712102K}.
\cite{2012NatCo...3E1315M}, using high resolution H$\alpha$ data, found an upper limit for the filling factor of chromospheric mottles, which are structures considered to be on-disk counterparts of type I spicules, of $\sim4-5 \%$.

\subsection{Visual discussion}

Figures 4 through 7 depict selected regions of interest (ROIs) from the time-series associated with Fig.~\ref{movie1} described in Section~\ref{analysis}. They display contours around the 304 detections, red features where RREs were detected and blue features for RBE detections, with the intensity being proportional to the contrast of such detections. The time-stamps and the coordinates are in the same parameter space as the time-series associated with Fig.~\ref{movie1}.

\begin{figure*}[!htb]
  \centering
\sbox0{\includegraphics{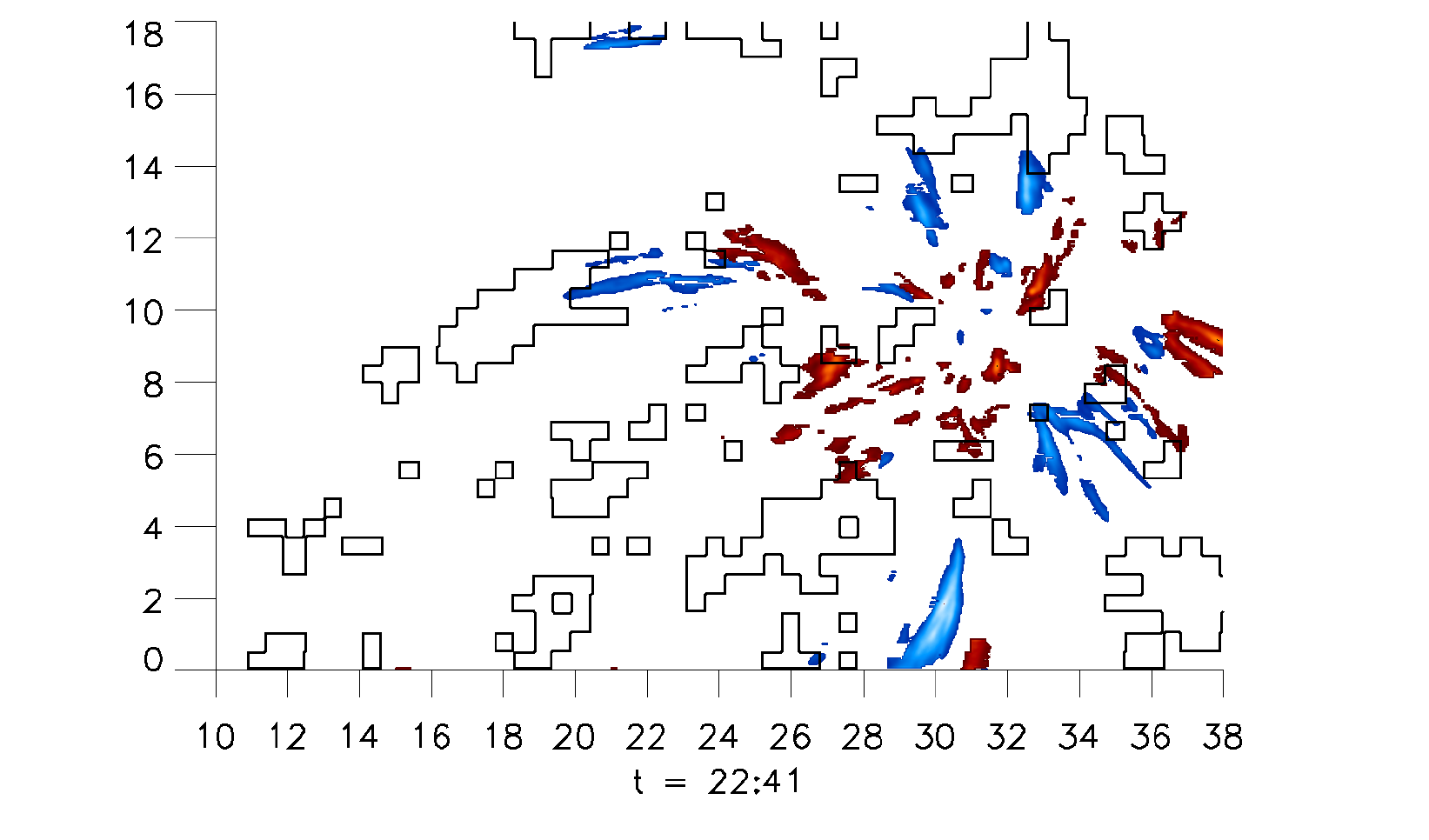}}%
\makeatletter
\Gscale@div\myscale{8cm}{\wd0}
\includegraphics[clip=true,trim={0.05\wd0} {0.02\wd0} {0.05\wd0} {0.05\wd0} ,scale=\myscale]{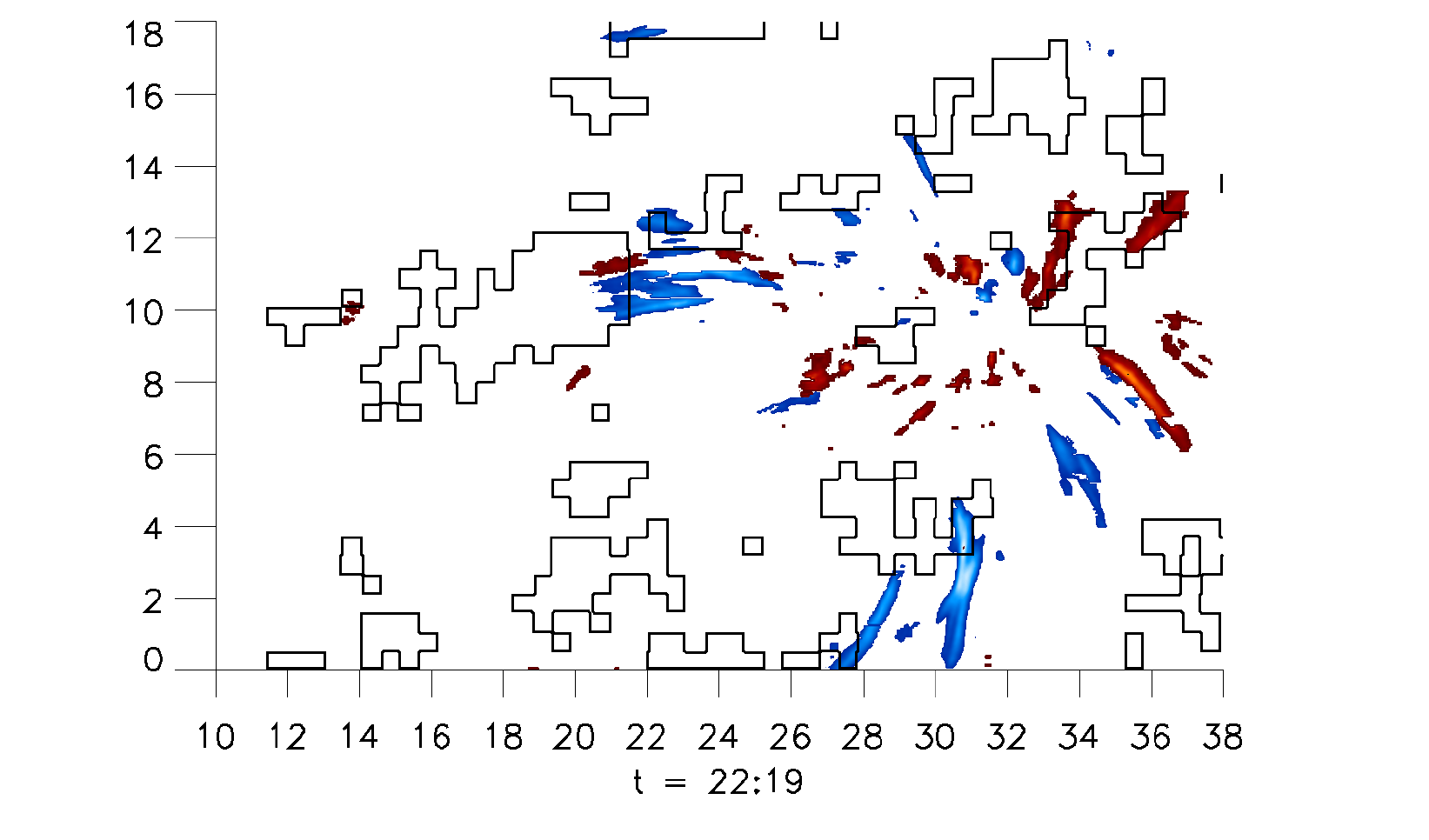}%shotgun
\includegraphics[clip=true,trim={0.17\wd0} {0.02\wd0} {0.05\wd0} {0.05\wd0} ,scale=\myscale]{00081-eps-converted-to.pdf}%shotgun
\includegraphics[clip=true,trim={0.17\wd0} {0.02\wd0} {0.05\wd0} {0.05\wd0} ,scale=\myscale]{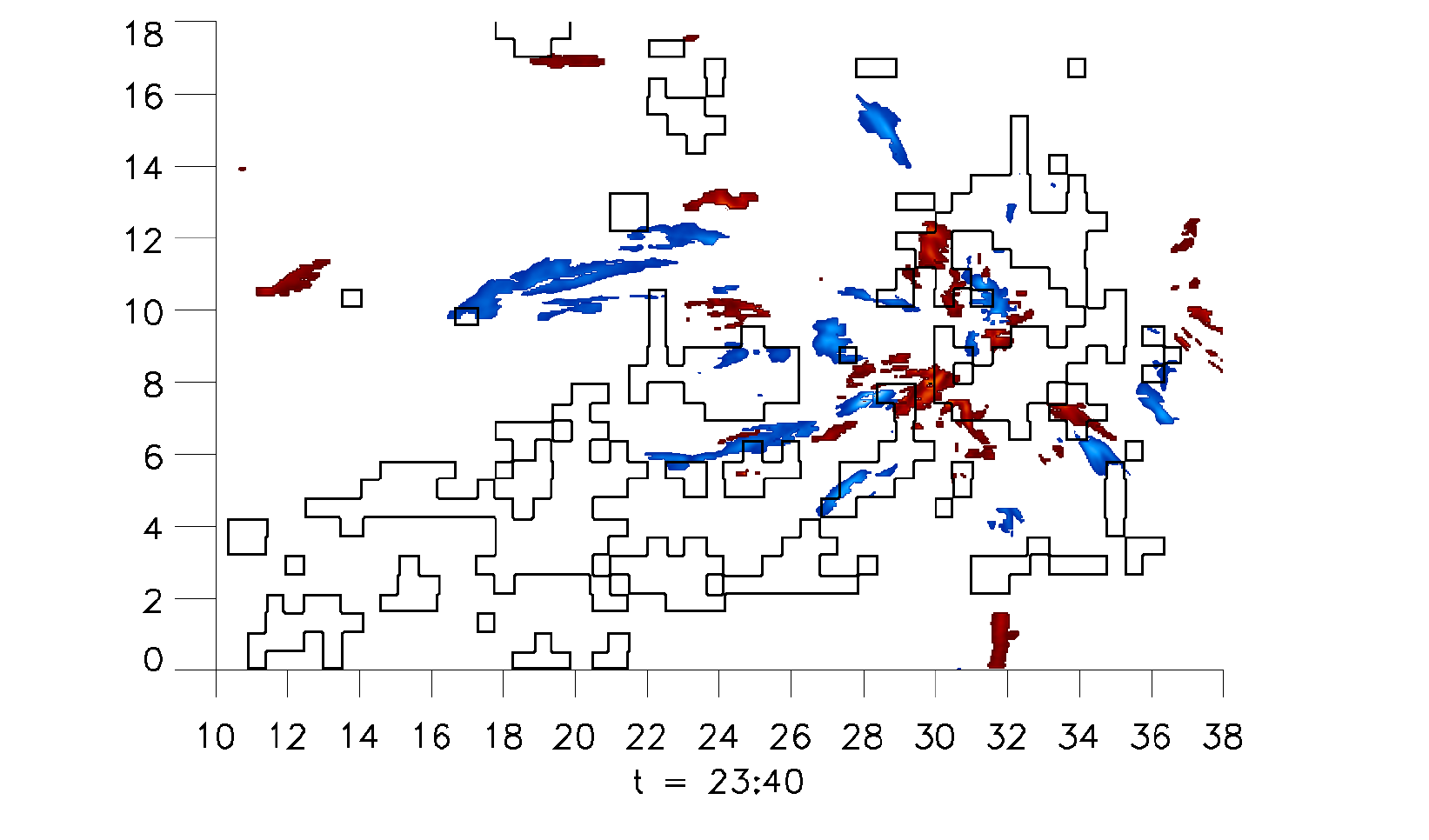}%shotgun
  \caption{\footnotesize  Lower rosette.  A large detection in 304 shown matching a stream of H$\alpha$ detections. This event is quasi-periodic. Red depicts RREs and blue RBEs. Tick-mark labels in terms of coordinates of the whole FOV in \arcsec.  Associated time-series available online. Time in mm:ss (from 9:00~AM).}
\label{lower}
\end{figure*}

Visually, the best matches between the H$\alpha$ features and 304 seem to be at the extremities of structures. Some matches, such as the one shown in Fig.~\ref{lower} at coordinates x=21\arcsec and y=10\arcsec, are very striking with a progression of REs into a large co-evolving 304 plume. However, these visual matches have nearly no overlap and many were missed by our overlap criteria. It is likely that the 11\% matching value would be increased if we match borders of 304 regions with the tips or the H$\alpha$ detections instead of matching overlaps over a large area. Such work, a computer vision problem, is being undertaken but the simple overlaps presented in this paper are significant and might indicate areas of multi-thermal gas or simply transition areas where the contribution for the opacity of H$\alpha$ and 304 intensity happens to be significant. (Note that the formation of H$\alpha$ fibrils is still largely unexplained, including formation temperatures). Similarly, we see more structures visually matched but not overlapping for RBEs than RREs. This observation may indicate that RREs tend to have a more multi-thermal structure whereas RBEs may become hotter "sooner" in the jet progression, which would then lead to higher overlaps between 304 and RREs but more RBEs disappearing into 304 detections.   However, the level of detections of RBEs and RREs are comparable in all aspects, with a slightly lower match of 304 to RREs (2 percentage points lower) but also slightly fewer RREs detected (leading to one percentage point lower matches expected due to randomness alone).

% ROSETTES, PLUMES, NON MATCHING 304
Multiple 304 brightenings as well as REs are visible quasi-periodically above and in the vicinity of the rosettes. Such is the case for the ROIs shown in Figures \ref{lower} and \ref{upper}. The ROI in Fig.\ref{upper} shows the last of a series of four pulses with only the latter showing RE counterparts. These counterparts, however, have very high contrast and are centered inside the 304 contours. The depicted pulse appears to have two lobes, with the lower one enveloping an RRE and the upper an RBE. Both the RRE and RBE are less than 0.6\arcsec wide and would have been missed in lower resolution studies.  As visible in the first 3 frames,  the signatures in 304 appear before the REs. We observe several instances of this effect, even when the H$\alpha$ feature seems to be propagating towards the 304 contour. This, together with the quasi-periodic nature of the four 304 pulses, matched with the occasional RE counterpart, at the same location and with the same propagating direction, suggests that the REs and the 304 brightenings have a common generating mechanism. Furthermore, this mechanism may generate jets at temperatures higher than the H$\alpha$ formation ranges at chromospheric heights.

\begin{figure*}[!htb]
  \centering
\sbox0{\includegraphics{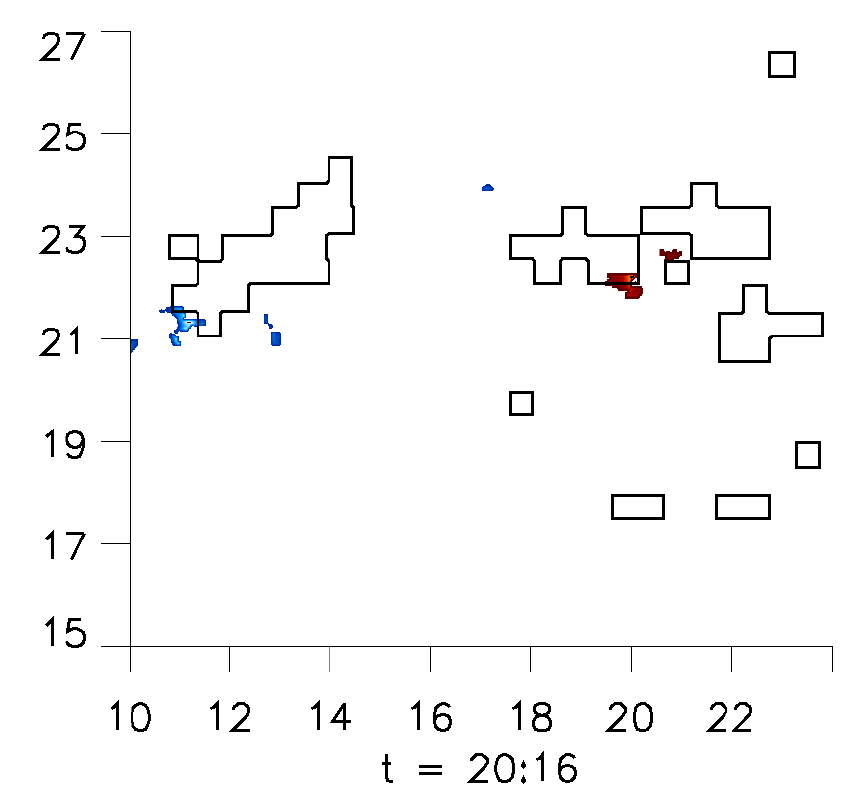}}%
\makeatletter
\Gscale@div\myscale{3.5cm}{\wd0}
\includegraphics[clip=true,trim=0 0 0 0,scale=\myscale]{00068-eps-converted-to.pdf}%norosette
\includegraphics[clip=true,trim={0.15\wd0} 0 0 0,scale=\myscale]{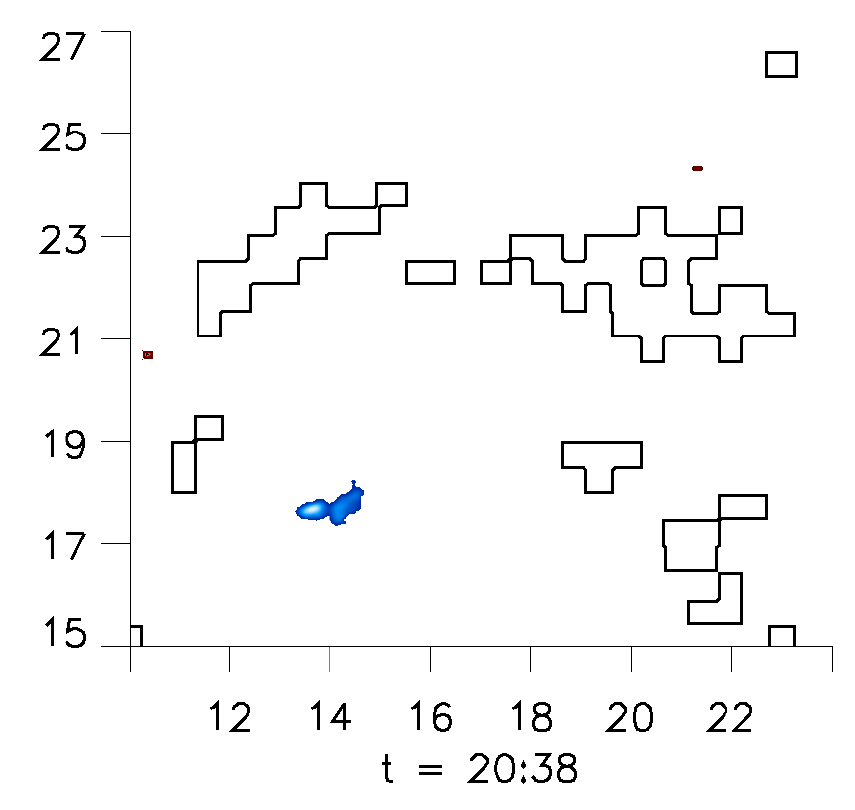}%norosette
\includegraphics[clip=true,trim={0.15\wd0} 0 0 0,scale=\myscale]{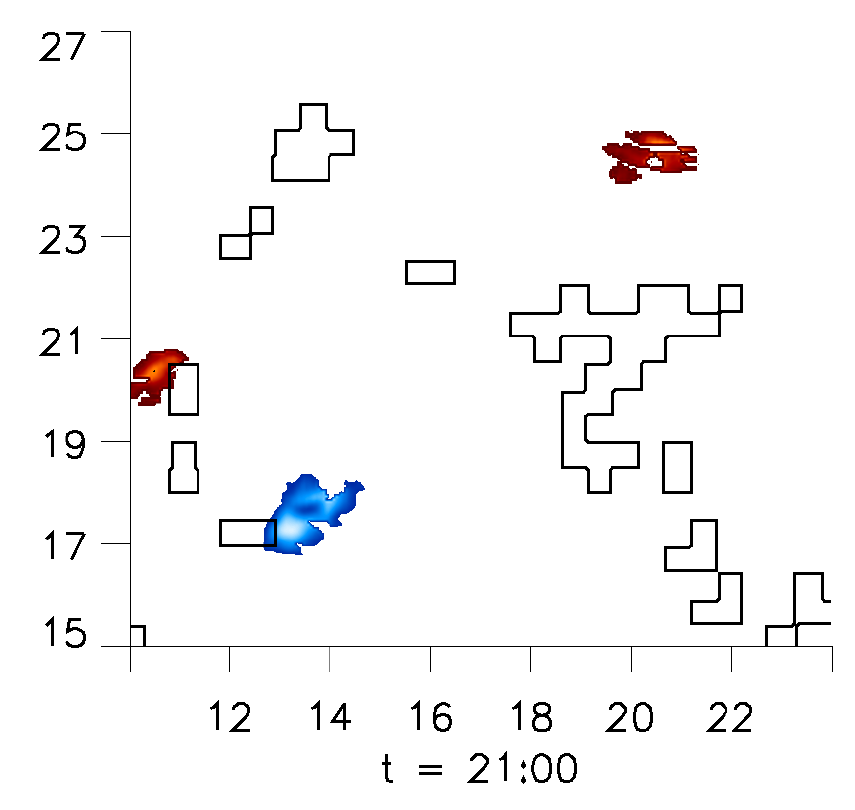}%norosette
\includegraphics[clip=true,trim={0.15\wd0} 0 0 0,scale=\myscale]{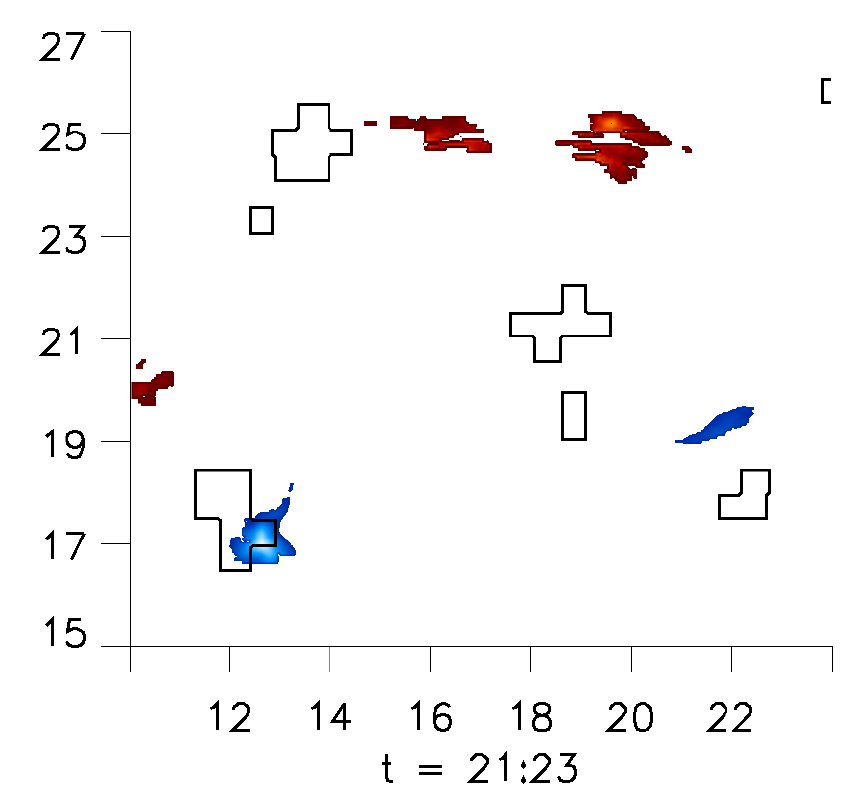}%norosette
\includegraphics[clip=true,trim={0.15\wd0} 0 0 0,scale=\myscale]{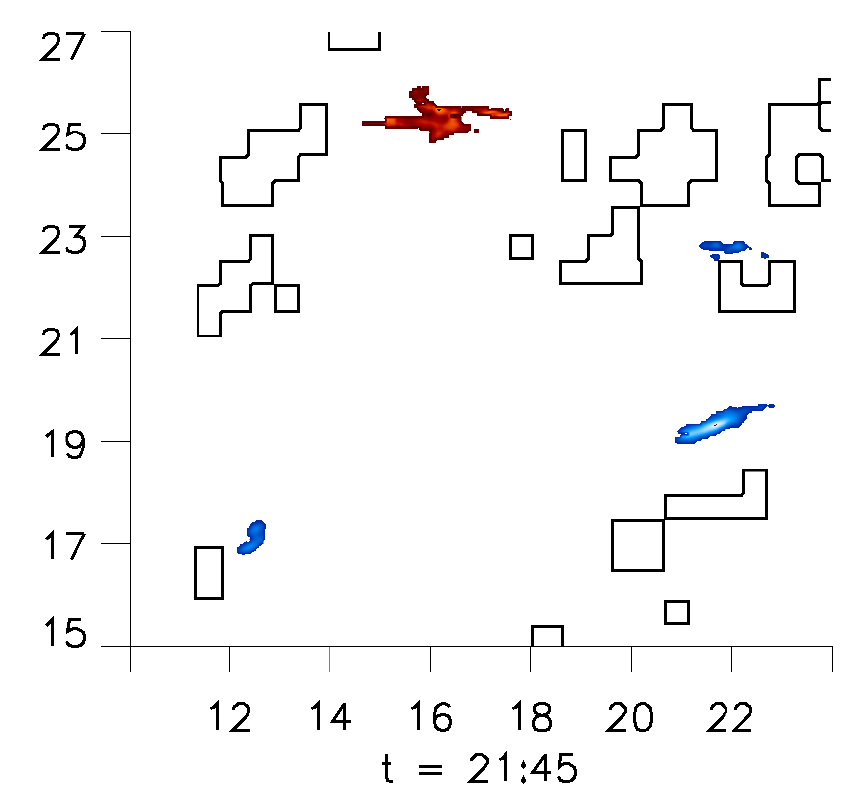}%norosette
\includegraphics[clip=true,trim={0.15\wd0} 0 0 0,scale=\myscale]{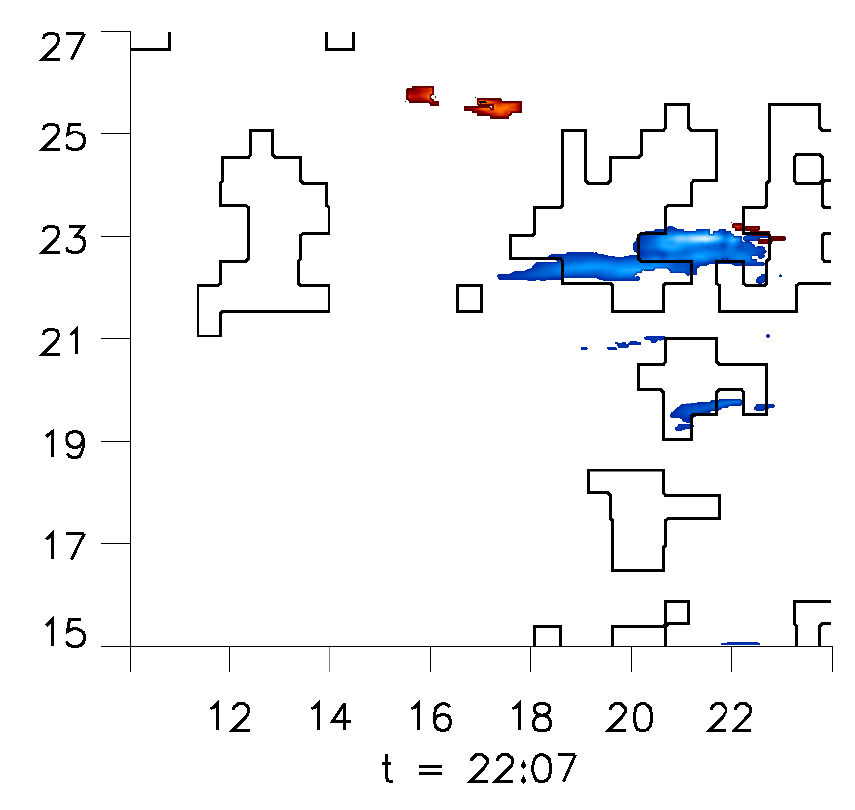}%norosette

\includegraphics[clip=true,trim=0 0 0 0,scale=\myscale]{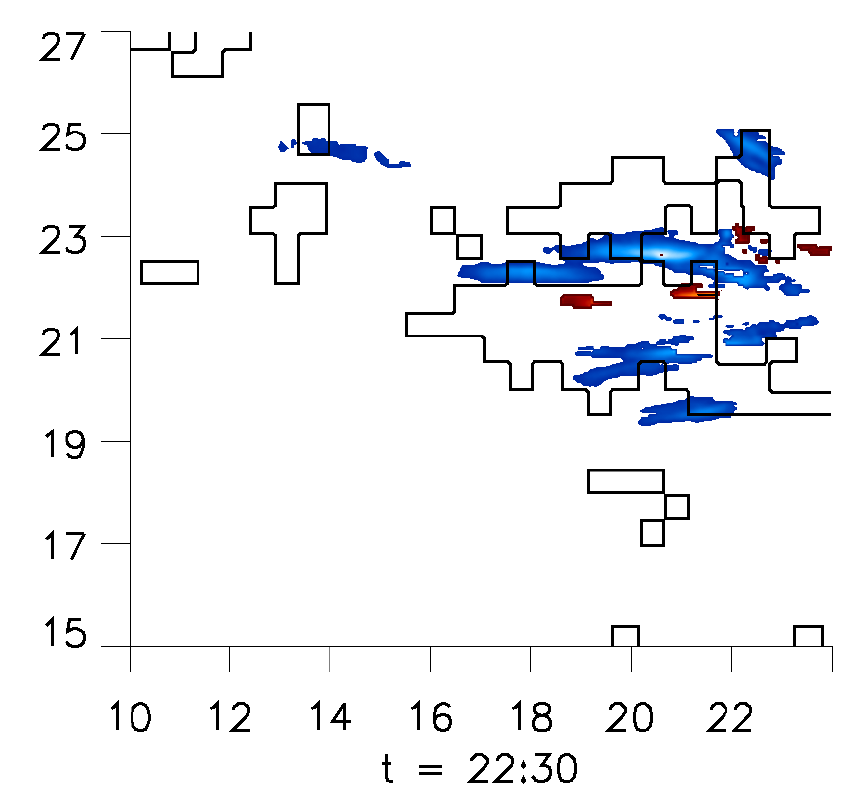}%norosette
\includegraphics[clip=true,trim={0.15\wd0} 0 0 0,scale=\myscale]{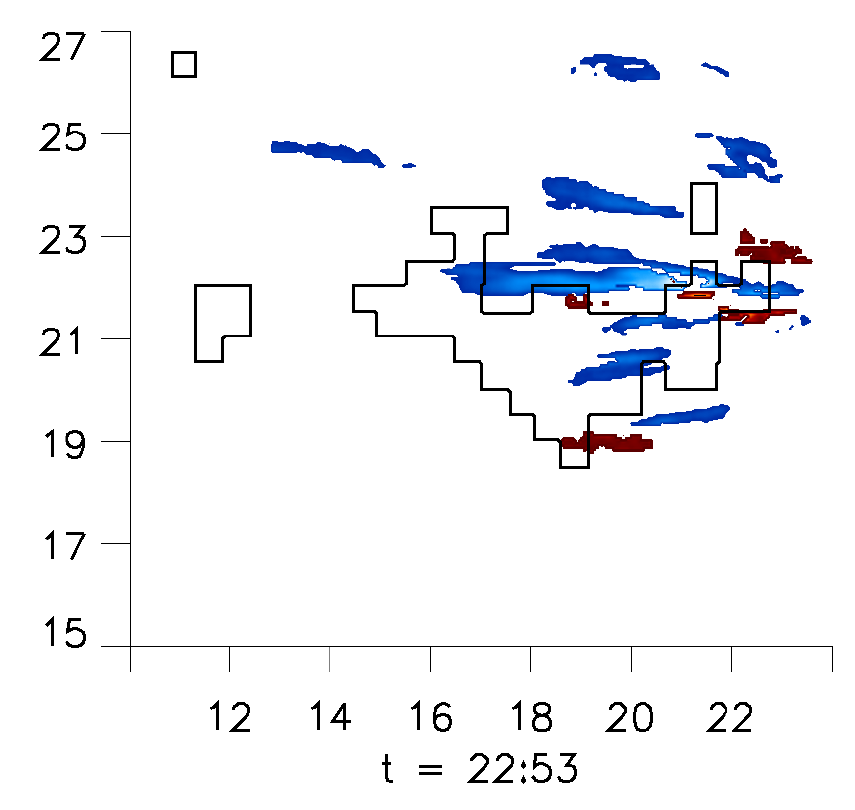}%norosette
\includegraphics[clip=true,trim={0.15\wd0} 0 0 0,scale=\myscale]{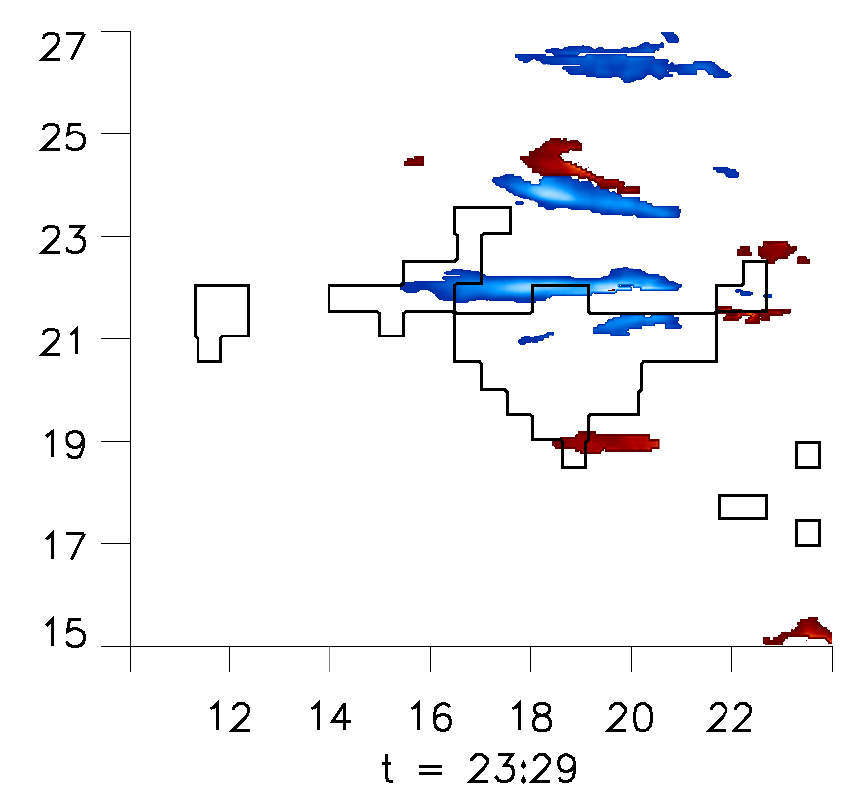}%norosette
\includegraphics[clip=true,trim={0.15\wd0} 0 0 0,scale=\myscale]{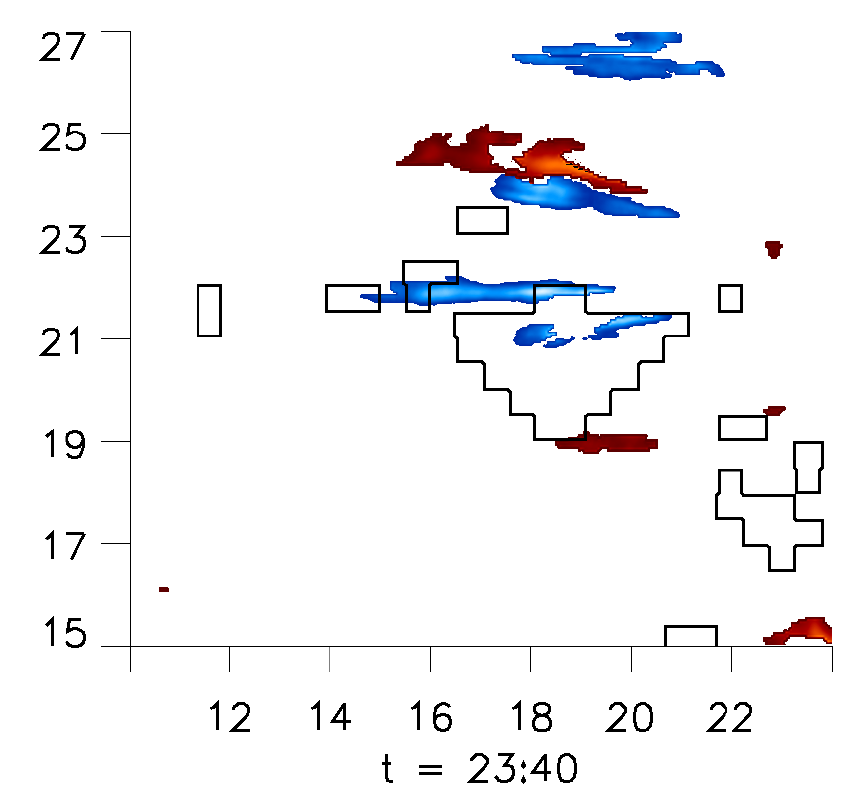}%norosette
\includegraphics[clip=true,trim={0.15\wd0} 0 0 0,scale=\myscale]{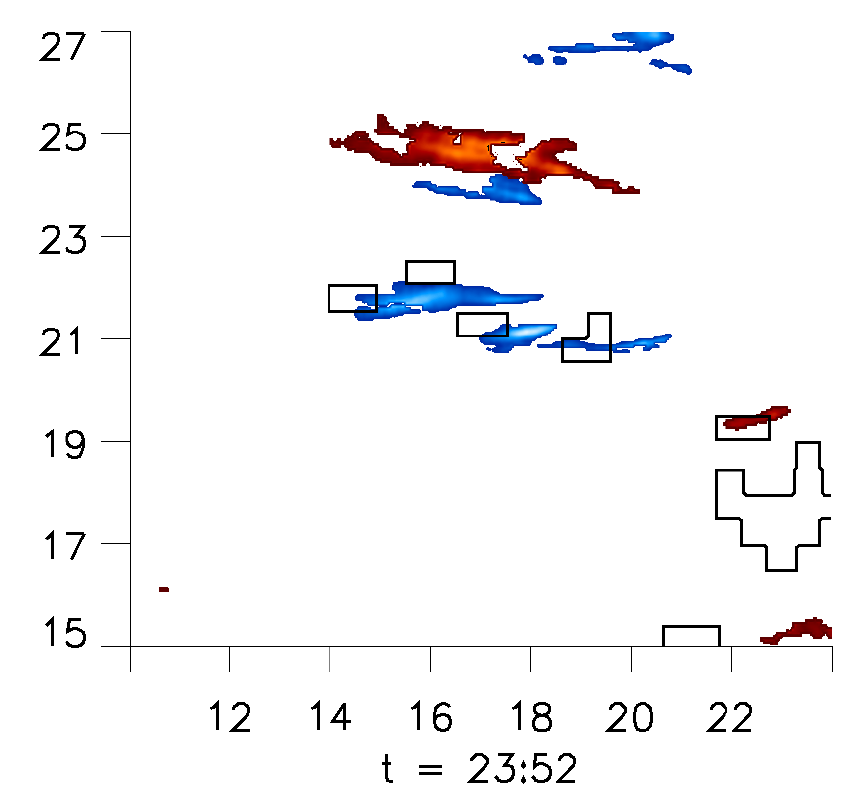}%norosette
\includegraphics[clip=true,trim={0.15\wd0} 0 0 0,scale=\myscale]{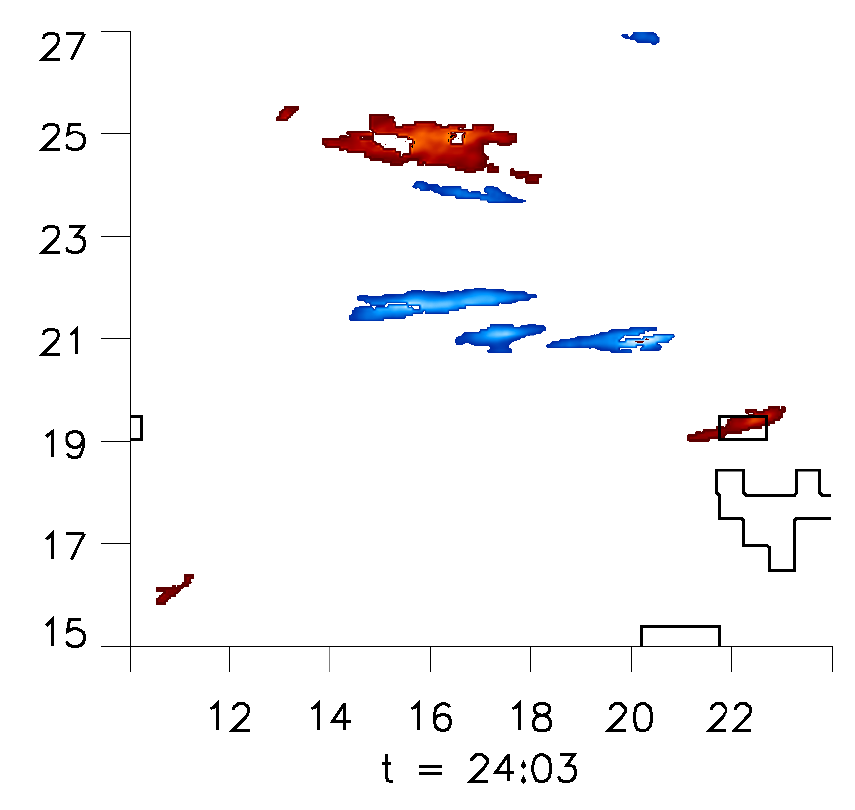}%norosette
  \caption{\footnotesize  ROI of a region away from network bright points (left). Red depicts RREs and blue RBEs. Units in arcseconds.  Associated time-series available online. Time in mm:ss (from 9:00~AM).}
\label{norosette}
\end{figure*}

% ULTRA-QUIET STRUCTURES
Brightenings also occur over areas that do not have any obvious relation with the rosettes nor any bright point concentrations. Figures \ref{norosette} and \ref{awayright} show three striking examples where 304 detections are matching RREs and RBEs. This type of events have not, as far as we are aware,  been detected before. The first row of Fig.~\ref{norosette} shows a small RBE blob with a 304 counterpart of similar dimensions at the bottom left corner. As may be seen from the intensity scale, the RBE has very high contrast and would have been detected even with very aggressive thresholding. It also has a very small elongation even though it has a well defined propagation direction. This example shows that any elongation criteria for RBE characterization would have removed features such as this. It also demonstrates that we are capturing very small-scale events and that these are important to the visible features of the lower corona.  The second row of Fig.\ref{norosette} shows a group of REs with an adjacent and slight overlapping co-evolving 304 detection. In the second to final frame, 3 small 304 signals are visible at the tips of the central RBEs.  

For the ROI in Fig.~\ref{awayright}, the match between H$\alpha$ and 304 occurs at the final location of the H$\alpha$ signature. The H$\alpha$ structure in Fig.~\ref{awayright}, in the time-series, has the appearance of an arch as seen from above, and the composing features of the arch appear to converge from the extremities to the centre. In the centre of the ''arch´´, at the end of the lifetime of the structure, an overlapping 304 signature is visible. This structure does not look like a jet but has spectral properties similar to those of an RBE as well as the transverse movements observed by \cite{2015ApJ...802...26K}. Note that the AIA detected region is only 3 AIA pixels wide and, with the time slices, barely large enough in our detection criteria to be counted as a match. 
We believe that such features, located in unremarkable locations away from any network elements, would be ubiquitous in the Sun.

\subsection{Visually matching 171 events to REs and 304}

The 171 detection map is noisier than the others. In this map we  focus on the most powerful events, and we observe multiple examples of matches across 304, 171 and REs. An example of this is clear in Fig.\ref{movie1} at the coordinates x=40\arcsec y=40\arcsec as well as in the associated time-sequence. The detections are elongated across all channels, along a diagonal in a very isolated region, and the 304 detection seems to appear first at frame t=08:44 followed by the RBE and the 171 signal that appear simultaneously. In the same location but at frame t=09:07, we have again a detection in 304 and 171 but without an RBE counterpart. Rather, this event seems to have a faint RRE counterpart delayed in time that we tentatively attribute to coronal rain. 

\begin{figure*}[!htb]
  \centering
%\resizebox{3cm}{!}{\includegraphics[clip=true]{f/f_white_version/cut0/00126-eps-converted-to.pdf}}
%\resizebox{3cm}{!}{\includegraphics[clip=true]{f/f_white_version/cut0/00127-eps-converted-to.pdf}}
\sbox0{\includegraphics{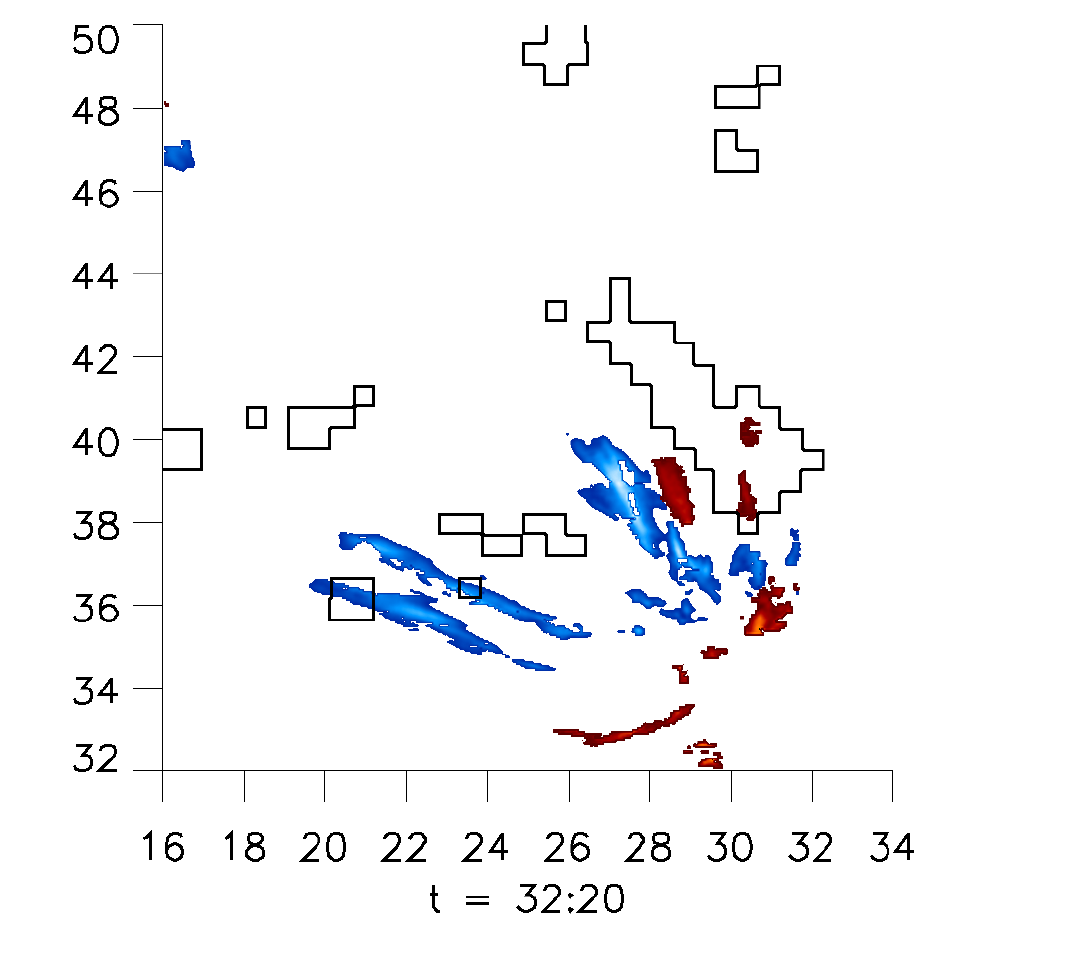}}%
\makeatletter
\Gscale@div\myscale{3.5cm}{\wd0}
\includegraphics[clip=true,trim=0 0 0 0,scale=\myscale]{00128-eps-converted-to.pdf}%norosette
\includegraphics[clip=true,trim={0.15\wd0} 0 0 0,scale=\myscale]{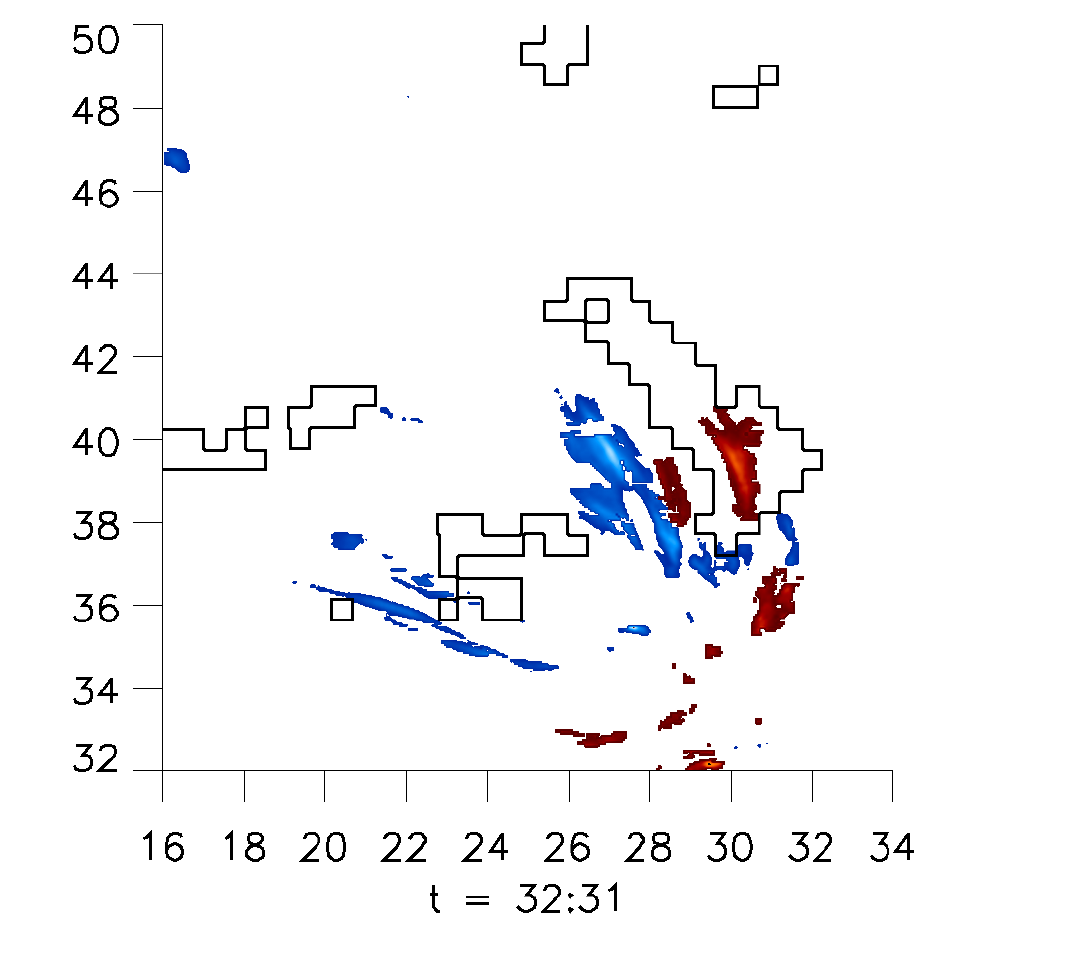}%norosette
\includegraphics[clip=true,trim={0.15\wd0} 0 0 0,scale=\myscale]{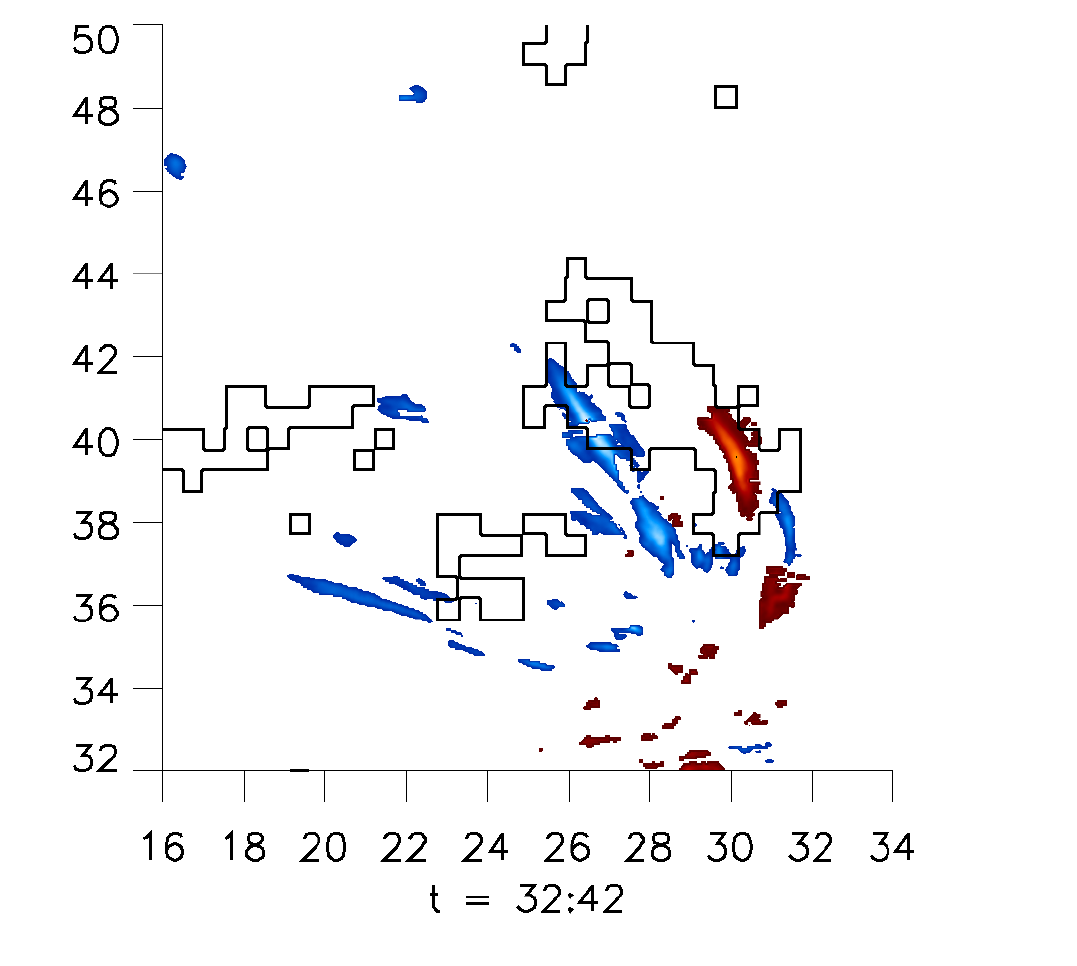}%norosette
\includegraphics[clip=true,trim={0.15\wd0} 0 0 0,scale=\myscale]{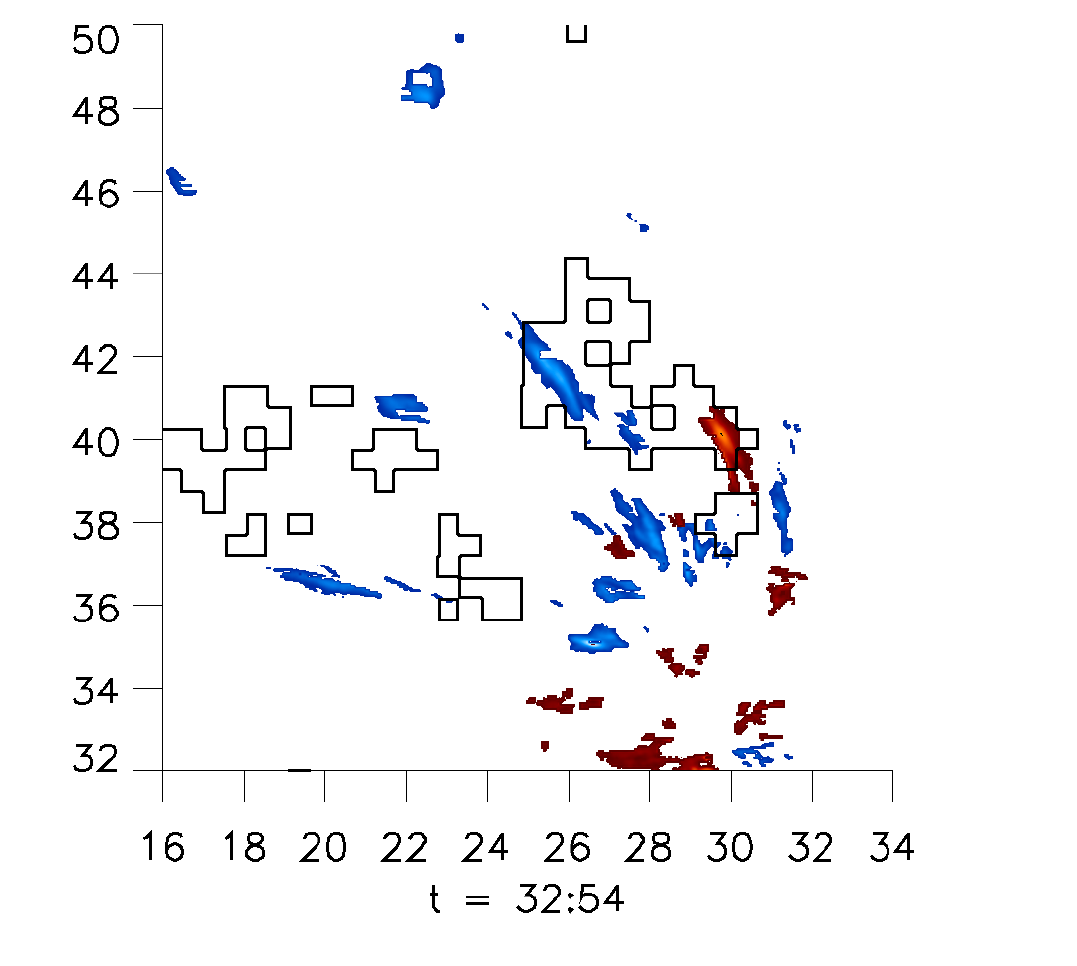}%norosette
\includegraphics[clip=true,trim={0.15\wd0} 0 0 0,scale=\myscale]{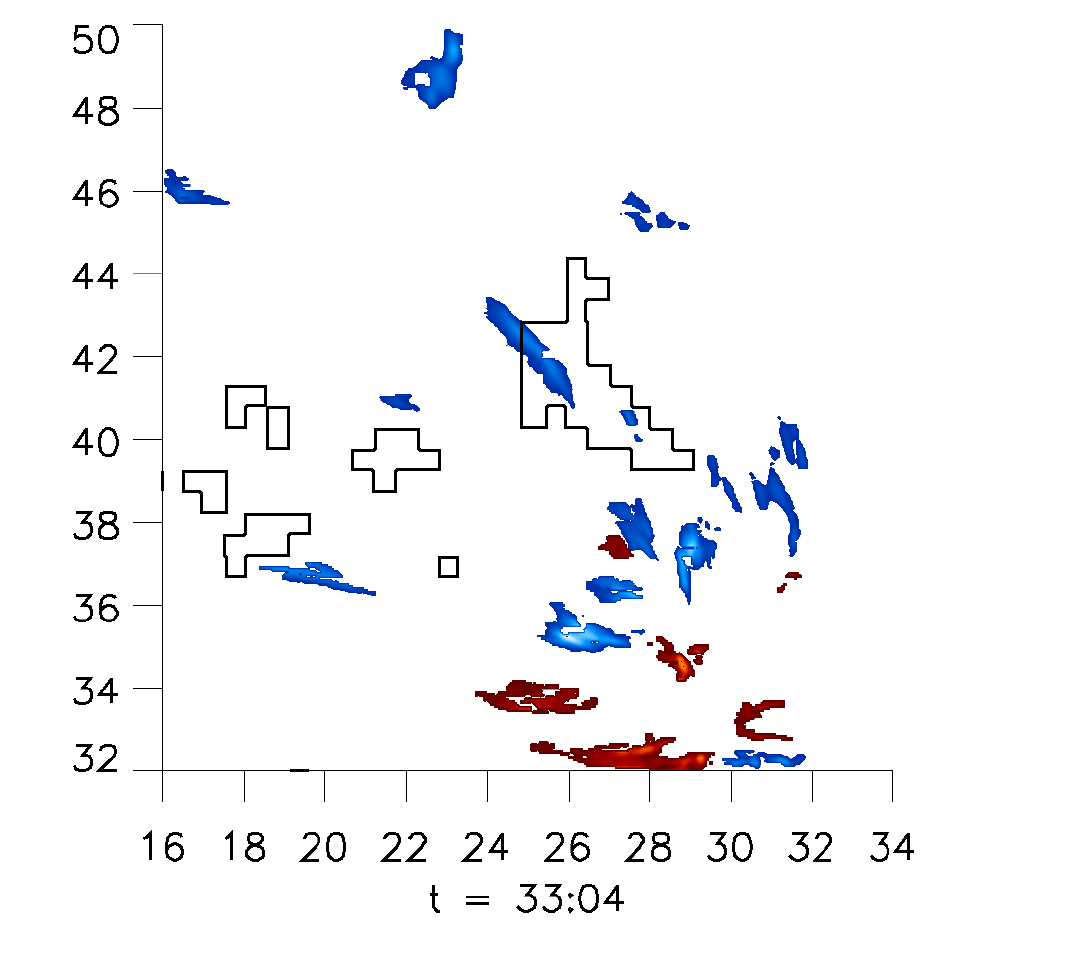}%norosette
\includegraphics[clip=true,trim={0.15\wd0} 0 0 0,scale=\myscale]{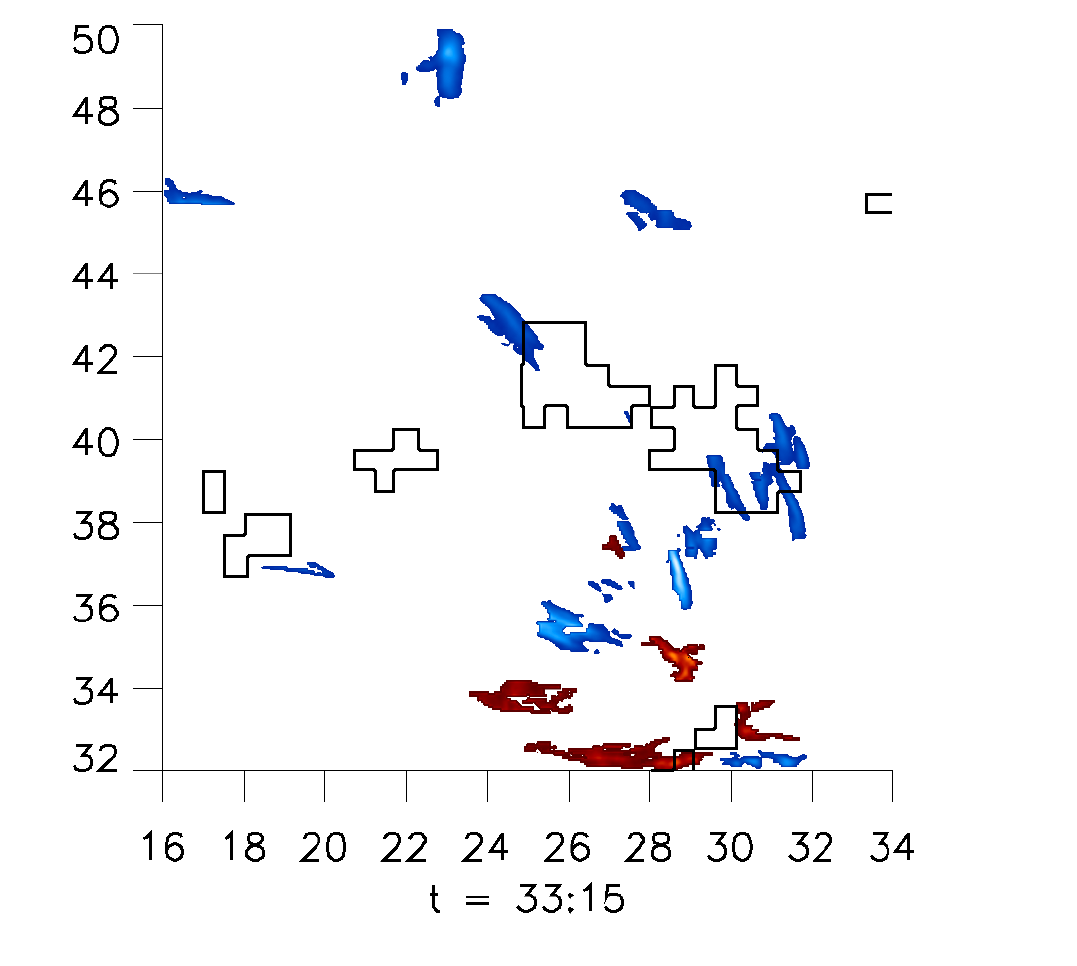}%norosette
  \caption{\footnotesize   Upper left region of the top rosette. Red depicts RREs and blue RBEs. Units in arcseconds.  Associated time-series available online. Time in mm:ss (from 9:00~AM).}
\label{upper}
\end{figure*}

\begin{figure*}[!htb]
  \centering
\sbox0{\includegraphics{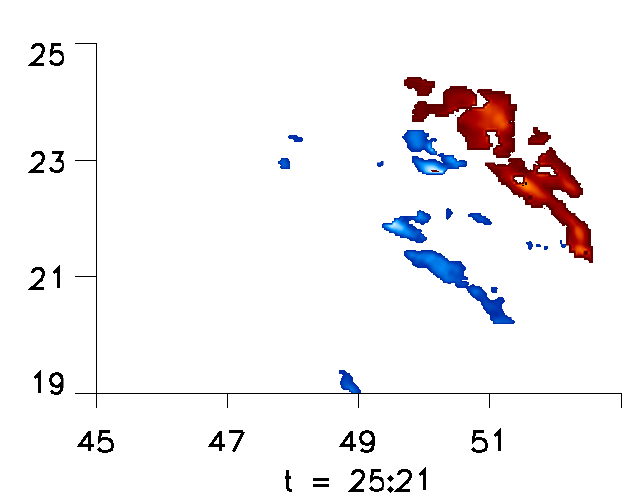}}%
\makeatletter
\Gscale@div\myscale{3.5cm}{\wd0}
\includegraphics[clip=true,trim=0 0 0 0,scale=\myscale]{00093-eps-converted-to.pdf}%norosette
\includegraphics[clip=true,trim={0.15\wd0} 0 0 0,scale=\myscale]{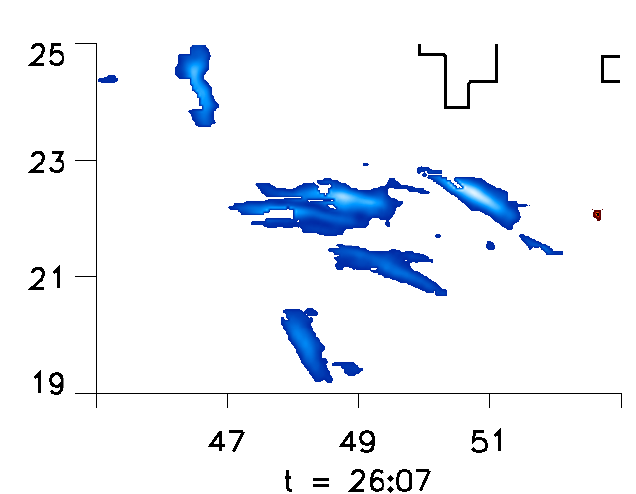}%norosette
\includegraphics[clip=true,trim={0.15\wd0} 0 0 0,scale=\myscale]{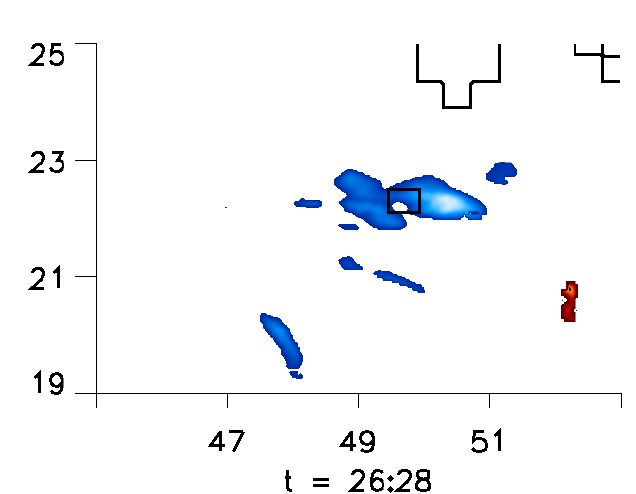}%norosette
\includegraphics[clip=true,trim={0.15\wd0} 0 0 0,scale=\myscale]{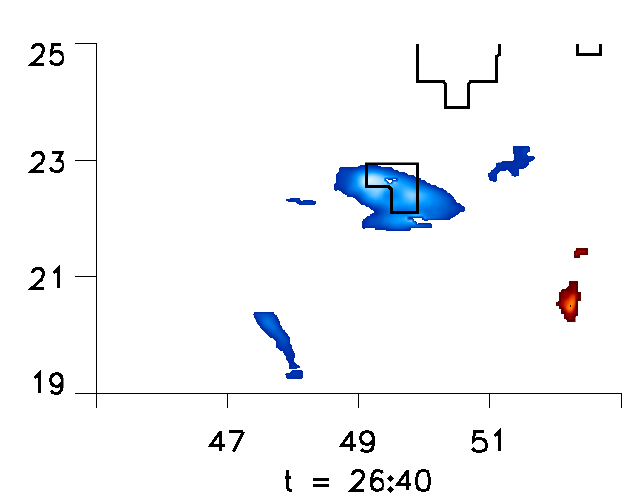}%norosette
\includegraphics[clip=true,trim={0.15\wd0} 0 0 0,scale=\myscale]{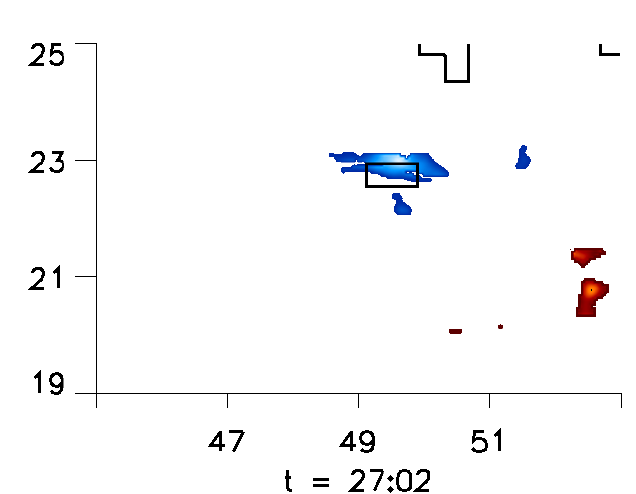}%norosette
\includegraphics[clip=true,trim={0.15\wd0} 0 0 0,scale=\myscale]{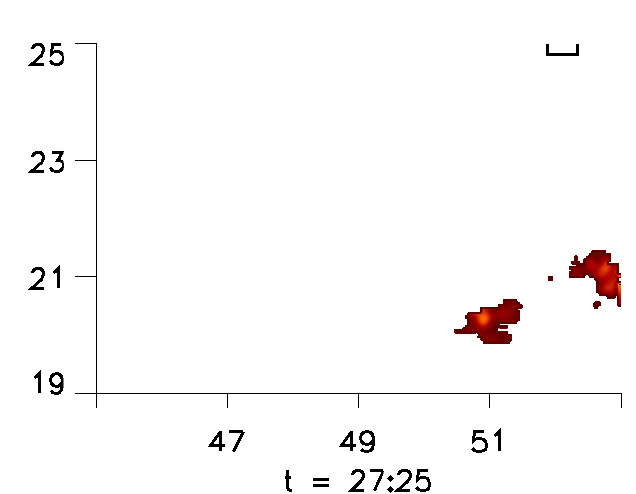}%norosette
  \caption{\footnotesize  Another region away from any bright-points and rosettes. Red depicts RREs and blue RBEs. Units in arcseconds.  Associated time-series available online. Time in mm:ss (from 9:00~AM). }
\label{awayright}
\end{figure*}

Examining the centre of the rosettes we can find more examples due to the quasi-periodic stream of detections across all channels. One such example is visible in the lower rosette at t=9:19 of the time-series associated with Fig.\ref{movie1} (coordinates x=30'' y=10'') which would correspond to looking straight-down at the centre of the fanning canopy. This can easily  be interpreted as the jet structure, directed at the observer, being visible at different heights as it is heated to coronal temperatures. Around the same timeframe one can see propagating events in the canopy, in all maps, at coordinates of x=20'', y=8''. At this location, detections in all channels seem to repeat throughout the time-series. Together with quasi-periodic pulses shown in Fig.~\ref{upper} and discussed in Section~\ref{discussion}, these multi-channel detections with a periodicity amount to evidence that jets exist that are generated by the same mechanisms but at a wide range of temperatures, including coronal temperatures, at very low geometrical heights.

% We also get the impression thar stornger events are more associated with slimer REs, maybe they disappear from H$\alpha$ due to heating?

%fig 4

% trim left lower right upper
%fig 5
% Tick-mark labels in terms of coordinates of the whole FOV in \arcsec

%fig 6
%
%
%
%fig 7

\section{Concluding remarks}

We confirm statistically, for the first time, that there is a relation between strongly Doppler-shifted H$\alpha$ transients at small scales in the quiet sun and heating signatures in the TR and corona. Noise limits our measured match-levels by design, limited by statistic significance, thus our values should be taken as lower bounds. It should be noted that a match implies overlapping in time as well as in space. Further, we find small features, matched in both H$\alpha$ and in the TR, that occur over very quiet sun without any obvious bright photospheric-footpoints. 

These two results show that  frequent small-scale transient-flows in the chromosphere, including but perhaps not limited to type II spicules, are indeed a significant contributor to the heating and dynamics of the upper atmospheric layers of the Sun including the corona. Although this has been discussed before, we show for the first time that these chromospheric transients are continuously reaching the upper layers of the solar atmosphere at very unremarkable regions of the Sun and provide values for the minimum amount of coronal events that can be traced back to chromospheric, quiet-Sun REs. 

We directly measure a filling factor of 1.3\% for RBEs and of 1\% for RREs in the quiet Sun, with about a quarter having a detectable signature in 304 and 171.

We further propose that REs and coronal transients often share a common source that generates repeated events but at different temperatures in each pulse. This is based on our direct observations of 304 signatures flowing out of a bright-point that sometimes, but not always, have a very clear H$\alpha$ RE counterpart. Such scenario would be consistent with jets of material being heated to low coronal temperatures at geometrical heights traditionally associated with the chromosphere, with some of these jets completely bypassing any meaningful intermediate signature at H$\alpha$ formation temperatures. 

%%%%%%%%%%%%%%%%%%%%%%%%%%%%%%%%%%%%%%%%%%%%%%%%%%%%%%%%%%%%%%

\begin{acknowledgements}
This work was first presented at the Hinode 9 conference, held at Queen's University Belfast in September 2015. We would like to thank Peter S\"{u}tterlin for assisting with the observations. We would also like to acknowledge support from Robert Ryans with computing infrastructure.    This work made use of an IDL port of Robert Shine's routines made by Tom Berger. Solarsoft was used. The Swedish 1-m Solar Telescope is operated on the island of La Palma by the Institute for Solar Physics (ISP) of Stockholm University
in the Spanish Observatorio del Roque de los Muchachos of the Instituto de Astrof\'isica de Canarias. This research was supported by the SOLARNET project (www.solarnet-east.eu), funded by the European Commissions FP7 Capacities Program under the Grant Agreement 312495. This work has been supported by the UK Science and Technology Facilities Council (STFC). D.K. would like to acknowledge funding from the European Commissions Seventh Framework Programme (FP7/2007- 2013) under grant agreement No. 606862 (F-CHROMA).
\end{acknowledgements}

\end{document}